\documentclass[12pt,a4paper]{article}
\pdfoutput=1  
\usepackage{amssymb}
\usepackage{amsmath}
\usepackage{amsfonts}
\usepackage{enumerate}
\usepackage{float}
\usepackage{verbatim}
\usepackage{graphicx}
\usepackage{lscape}
\usepackage[autostyle]{csquotes}
\usepackage{bbm}
\usepackage[dvipsnames]{xcolor}
\usepackage{pdflscape,array,booktabs}
\usepackage[a4paper,bindingoffset=0.2in,left=1in,right=1in,top=1in,bottom=1in,footskip=.5in]{geometry}
\usepackage{caption}
\usepackage{siunitx,booktabs} 
\usepackage{hyperref}
\usepackage[normalem]{ulem}
\useunder{\uline}{\ul}{}
\usepackage[title]{appendix}
\usepackage{tabularray}

\usepackage{siunitx}
\usepackage{changepage}

\graphicspath{ {images/} }   

\DeclareMathOperator*{\argmin}{arg\,min}
\DeclareMathAlphabet{\pazocal}{OMS}{zplm}{m}{n}

\usepackage[backend=biber, maxbibnames=99, maxcitenames=2, uniquelist=false, natbib=true, bibstyle=apa, citestyle=authoryear,  isbn=false, uniquename=false,doi =false, url=false]{biblatex}
\AtEveryBibitem{%
  \clearfield{note}%
}

\begin{document}

\title{Human Wellbeing and Machine Learning}
\author{  Ekaterina Oparina\footnote{These authors are joint first authors. The displayed order of these authors is random and was determined by the AEA randomization tool, confirmation code $oJsh\_ZMZJwhH$. Author affiliations: Ekaterina Oparina: London School of Economics; Niccolò Gentile, Conchita D'Ambrosio and Alexandre Tkatchenko: University of Luxembourg; Caspar Kaiser and Jan-Emmanuel De Neve: University of Oxford;  Andrew E. Clark: Paris School of Economics - CNRS. \newline We thank Filippo Volpin for excellent research assistance and Sid Bhushan for early discussions on the topic. We are grateful to the participants of the LSE Wellbeing Seminar for their comments and suggestions. Funding via the ERC Grant Agreement n. 856455, and the Institute for Advanced Studies, University of Luxembourg, Grant DSEWELL is gratefully acknowledged. We thank The Gallup Organization for providing access to their data for this research project.}  \and \textcircled{r} \and   Caspar Kaiser$^*$  \and \textcircled{r} \and 
   Niccolò Gentile$^*$ \and   Alexandre Tkatchenko  \and \quad\quad Andrew E. Clark \and \quad Jan-Emmanuel De Neve  \and Conchita D'Ambrosio}

\date{\today}
\maketitle

\begin{abstract}
There is a vast literature on the determinants of subjective wellbeing. International organisations and statistical offices are now collecting such survey data at scale. However, standard regression models explain surprisingly little of the variation in wellbeing, limiting our ability to predict it. In response, we here assess the potential of Machine Learning (ML) to help us better understand wellbeing. We analyse wellbeing data on over a million respondents from Germany, the UK, and the United States. In terms of predictive power, our ML approaches do perform better than traditional models. Although the size of the improvement is small in absolute terms, it turns out to be substantial when compared to that of key variables like health. We moreover find that drastically expanding the set of explanatory variables doubles the predictive power of both OLS and the ML approaches on unseen data. The variables identified as important by our ML algorithms – \textit{i.e.} material conditions, health, and meaningful social relations – are similar to those that have already been identified in the literature. In that sense, our data-driven ML results validate the findings from conventional approaches.
\\
\\
\\

\end{abstract}


\vspace{1.5in}

\setcounter{page}{1}\newpage

\section{Introduction}

Over the last 40 years, researchers from various fields have established an immense literature on the correlates and determinants of subjective wellbeing (\citet{clark_four_2018}, \citet{diener_advances_2018}, \cite{nikolova_economics_2020}). In parallel, international organisations (\citet{oecd_hows_2020}) and national governments (\citet{ons_well-being_2021}) have turned to subjective wellbeing data as a key tool for policy analysis. 

However, despite the widespread use of wellbeing scores, our current ability to predict wellbeing is limited. Conventional linear models, where variables are selected based on intuition or theory, explain little individual-level variation. Typically, models of individual wellbeing produce an R-squared of no more than 15\%.  

In response, we here evaluate whether machine learning (ML) algorithms can improve our capacity to understand wellbeing. We answer two research questions:\\
\begin{adjustwidth}{20pt}{10pt} 
\textbf{RQ1:} Are ML algorithms significantly better at predicting wellbeing than conventional linear models? What is the upper limit on our ability to predict wellbeing based on survey data?\\
\textbf{RQ2:} Are the variables that are identified by ML algorithms as important in predicting wellbeing the same as those in the conventional literature? \\
\end{adjustwidth}

To answer these questions, we use random forests (\citet{breiman_random_2001}, \citet{hastie_elements_2009}), gradient boosting (\citet{friedman_greedy_2001}, \citet{natekin_gradient_2013}), and penalised regressions (\citet{tibshirani_regression_1996}) as examples of ML algorithms. Random forests and gradient boosting are tree-based algorithms that have been shown to perform well with tabular data (\citet{shwartz-ziv_tabular_2022}).\footnote{In contrast, other ML algorithms, such as neural networks, tend to perform poorly on tabular data, which is why we do not consider them here (\citet{borisov_deep_2022}). In preliminary analyses we did indeed find that feed-forward neural networks yielded performances that were no better than OLS.} Penalised regressions are a convenient tool for analyses that involve large number of covariates, like ours (\citet{tibshirani_regression_1996}).
Generally, these techniques can identify more-complex models of wellbeing than traditional linear models, potentially improving predictive performance. Unlike standard regression techniques, these algorithms allow for the inclusion of an arbitrary number of variables, and, in the case of our tree-based methods, can identify nonlinearities and interactions between variables.

To the best of our knowledge, this paper is the first systematic attempt to evaluate the (dis) advantages of using ML for studying wellbeing at a global scale. Earlier work focused on relatively small country- and year-specific samples (\citet{margolis_what_2021}), or particular drivers of wellbeing, such as age (\citet{kaiser_using_2022-1}).  

We carry out our empirical analysis using three of the largest currently-available datasets that include wellbeing information: the German Socio-Economic Panel (SOEP), the UK Household Longitudinal Study (UKHLS), and the American Gallup Daily Poll. The SOEP has data on about 30,000 unique respondents and 400 distinct variables; the UKHLS surveys around 40,000 individuals in each wave and has over 500 distinct variables; and each year of the Gallup data has information on around 200,000 respondents with approximately 60 distinct variables. We can thus study the extent to which utilising more information about  individual respondents improves the predictive power of wellbeing models. 

Regarding RQ1, we find that ML algorithms predict somewhat better than standard linear models. The size of this improvement is small in absolute terms, but substantial when compared to the predictive power of key variables, such as health. Increasing the number of variables in the model from a standard set (we call this the \enquote{Restricted Set}) to all available data (the \enquote{Extended Set}) has a far larger effect on predictive model performance. Predictive accuracy, judged by the R-squared on unseen data, roughly doubles for both OLS and ML methods. Independently of the type of algorithm, an R-squared of 0.30 appears to be the feasible maximum given the available data.\footnote{Our estimations on the extended set of variables, which include all of the variables apart from other measures of subjective wellbeing, produce R-squared figures of between 0.25 and 0.32 across the different datasets.} 

For RQ2, our data-driven ML results validate the findings of the conventional literature. We find that variables related to respondents’ social connections, health and material conditions are consistently among the most important in predicting wellbeing. Variable importance is assessed using permutation importances (\citet{breiman_random_2001}, \citet{kuh_mortality_2002}) and by computing pseudo partial effects for all algorithms, including OLS. In general, there is substantial correlation in variable importance rankings across algorithms ($\rho=0.58$ to $\rho=0.83$), so that ML approaches and OLS are largely in agreement in terms of what matters most for wellbeing.  

\section{Materials and Methods}
\subsection{Data}

We analyse data from three nationally-representative surveys over the 2010 to 2018 period: the German Socio-Economic Panel (SOEP), the UK Longitudinal Household Survey (UKHLS), and the US Gallup Daily Poll (Gallup). 

The Gallup data covers the US adult population, with daily cross-sectional telephone-based surveys of 500 (1000 until 2012) respondents. After removing incomplete data, this yields an annual sample ranging from N=115,192 (in 2018) to N=351,875 (in 2011). Wellbeing is measured by the Cantril Ladder of Life (\citet{cantril_pattern_1965}), which asks: \textit{Please imagine a ladder with steps numbered from zero at the bottom to ten at the top. The top of the ladder represents the best possible life for you and the bottom of the ladder represents the worst possible life for you. On which step of the ladder would you say you personally feel you stand at this time? }Answers are recorded on a scale from 0 to 10, with equal steps between response options.\footnote{There has been controversy about whether such data can support inferences about underlying wellbeing (\citet{bond_sad_2019}, \citet{chen_robust_2019}, \citet{kaiser_how_2020}, \citet{schroder_revisiting_2017}). We here remain agnostic about this issue. We instead rather ask which algorithms and models best predict the answers to wellbeing questions, without making any further claims about how these answers relate to respondents’ underlying feelings.} 

The SOEP is representative of the German adult population, with interviews conducted in person. To allow for a direct comparison with the Gallup data, we here consider the survey period between 2010 and 2018. In each year, between N=26,000 and N=30,000 observations are available. Life satisfaction is measured on a scale from 0 to 10, from the question: \textit{We would like to ask you about your satisfaction with your life in general, please answer according to the following scale: 0 means completely dissatisfied and 10 means completely satisfied: How satisfied are you with your life, all things considered? }

The UKHLS is representative of the UK adult population. Interviews are conducted in person. We again confine our analysis to the same 2010-2018 period (corresponding to Waves 2 to 10). The number of available annual observations is between N=29,605 to N=40,679. Life satisfaction is measured on a 1 to 7 scale. Respondents are asked: \textit{How dissatisfied or satisfied are you with your life overall?}. 

\begin{figure}[]
\caption{Histograms of life satisfaction for SOEP, UKHLS and Gallup data.}
\centering
\label{fig1}
\includegraphics[width=0.9\textwidth]{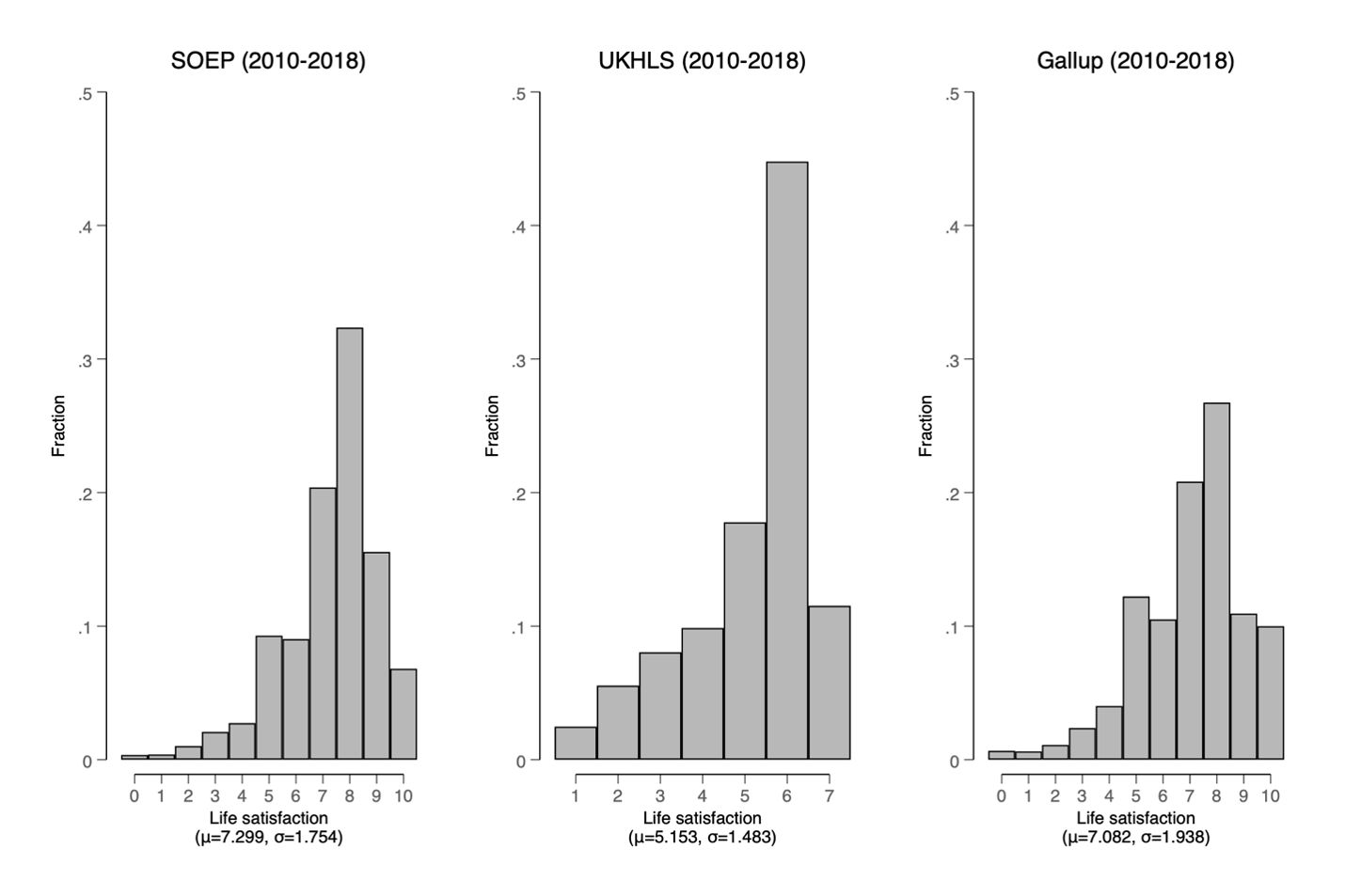}
\end{figure}

Descriptive statistics and histograms of each wellbeing measure appear in Figure \ref{fig1}. The wellbeing distributions are very similar across datasets. As is typically found in high-income countries, wellbeing is strongly left-skewed. 

\subsection{Algorithms}

We model wellbeing using four kinds of algorithms. 
First, as our baseline and corresponding to the workhorse of a great deal of research on subjective wellbeing, we apply \textbf{Ordinary Least Squares (OLS)} regressions. OLS estimates are the solution to the problem:
\begin{equation}
    \argmin_b\sum_{i=1}^N(x_i'b-s_i)^2
    \label{eq1}
\end{equation}

Here, $x_i$ is a vector of explanatory variables and $b$ the vector of coefficients. The wellbeing of respondent $i$ is denoted by $s_i$. Let $\hat{b}$ be the solution of Equation \ref{eq1}. Then, the predicted wellbeing level on the respondent $i$  is $\hat{s}_i = x_i'\hat{b}$. When using OLS, the researcher implicitly assumes that reported wellbeing is a linear combination of the chosen set of explanatory variables   $x$. If these assumptions are an appropriate description of the true data-generating process, OLS will provide accurate predictions of individual wellbeing. In applications with a large number of covariates, the performance of OLS may degrade due to overfitting or multicollinearity between included explanatory variables. 

The second algorithm we consider, the \textbf{Least Absolute Shrinkage and Selection Operator (LASSO)}, tackles this issue by adding a penalty for the sum of the magnitudes of the estimated coefficients. In particular, LASSO estimates are the solution to:
\begin{equation}
    \argmin_b\sum_{i=1}^N(x_i'b-s_i)^2+\lambda\sum_{k=1}^K|b_k|
\end{equation}
Here, $\lambda$ is a hyperparameter, the preferred value of which is found using a grid search. LASSO and OLS are equivalent for $\lambda=0$. Although LASSO may improve predictions by reducing the risk of overfitting, the algorithm continues to assume an additive functional form. Nevertheless, one helpful property of LASSO is that it shrinks coefficients on the variables with low explanatory power to zero. In some specifications, we thus use LASSO as a device for variable selection. 

The third and fourth algorithms we consider – \textbf{Random Forests (RF)} and \textbf{Gradient Boosting (GB)} – are based on regression trees (\citet{breiman_classification_1984}). Regression trees are generated via a recursive binary splitting algorithm. The algorithm splits the sample along values of covariates and predicts the outcome in each subsample, or \textit{node}, as the mean outcome within each node. More formally, at each step $k$, the data $D$ is split into two nodes $D_{L,k}$ and $D_{R,k}$. The location of the split within the data is determined by some variable $x_j$ and an associated threshold $\tau_{(k,j)}$. The nodes $D_{L,k}$ and $D_{R,k}$ are defined as (see \citet{hastie_elements_2009}):
\begin{equation}
   D_{(L,k)}= \{x | x_j < \tau_{k,j} \}; \\
   D_{(R,k)}= \{x | x_j \geq \tau_{k,j} \} 
\end{equation}

The predicted values are the mean value of $s$ within each node, \textit{i.e.} $\hat{s}_{D_{m,k}} = N^{-1}_{D_{m,k}}\sum_{i:X_i \in D_{m,k}}s_i$, for $m \in \{L,R\}$, where $N_{D_{m,k}}$ is the number of respondents in each node. The splitting variable $x_j$ and the thresholds $s_t$ are determined by minimising the following residual sum of squares:
\begin{equation}
   min_{j,s} \sum_{i:X_i \in D_{L,k}}(s_i-\hat{s}_{D_{L,k}})^2 + \sum_{i:X_i \in D_{R,k}}(s_i-\hat{s}_{D_{R,k}})^2
\end{equation}

Finally, the nodes $D_{L,k}$ and $D_{R,k}$ are in turn used as inputs for the next step. This procedure is repeated until some final number of \textit{leaves} is found. By construction, every split reduces the in-sample mean squared error (MSE).\footnote{Mean squared error measures the average of the squares of the errors – the average squared difference between the predicted and reported levels of wellbeing.} Hence, if the size of the tree is not limited, the algorithm will overfit the data. Limiting the maximum tree size can ameliorate this issue by reducing the variance of the predictions. However, this comes at the cost of increasing the bias of the resulting estimates (\citet{hastie_elements_2009}). Alternatively, the variance in the predictions can be reduced by aggregating the predictions from multiple trees. Random forests and gradient boosting are both examples of this strategy.

Specifically, \textbf{Random Forests}, the third algorithm we consider, rely on averaging across a large number of trees (which we set to 1,000 for all the three datasets).\footnote{The performance of the random forest is non-decreasing in the number of trees. In our application, increasing the number of trees to 2,000 for UKHLS and Gallup and to 10,000 for SOEP yields qualitatively similar results. The final number of trees was chosen to render the optimisation less computationally-expensive.}  Each individual tree has low bias but high variance. When the correlation between the trees is low, averaging across the predictions of multiple trees reduces the variance of the predictions without introducing additional bias. To carry out this procedure, each individual tree is grown on a nonparametrically bootstrapped sample of the original data. The correlation between trees is further reduced by considering only a random subset of all covariates at each split. The size of this subset, \textit{Nvars}, is a hyperparameter that we select based on a grid search.

The fourth algorithm, \textbf{Gradient Boosting}, proceeds by sequentially fitting regression trees on the residuals of the predictions of the previous collection of trees.\footnote{This construction of the trees, when the residuals from the previous tree are used to build the following tree, is specific to a case when the partitioning of the tree is chosen to minimise the sum of squared residuals in each node. The construction differs when other objective functions are used. See \citet{friedman_greedy_2001} and \citet{hastie_elements_2009} for the general case. We here use a standard implementation of gradient boosting. In preliminary tests, we also evaluated the performance of extreme gradient boosting (XGBoost; \citet{chen_xgboost_2016}) in the Gallup and SOEP datasets. The use of XGBoost only yielded negligible improvements compared to standard gradient boosting, which is why we here focus on the latter.}  Intuitively, each subsequent tree attempts to explain the variance that was not explained by the previous trees. We begin with the predictions $\hat{s}_{T{}_1}$  of a first tree $T_1$ and calculate the residual $\hat{s}_{T{}_1}-s_i = {e}_{T{}_1}$ . A second tree $T_2$ is then fitted on these residuals to obtain predicted residuals ${e}_{T_1}$. The updated overall predictions are then given by $\hat{s}_{T{}_1}+\hat{e}_{T{}_1} = \hat{s}_{T{}_2}$. A third tree is subsequently trained on the residuals $\hat{s}_{T{}_2}-s_i = {e}_{T{}_2}$. This process is repeated \textit{Ntrees} times, producing increasingly accurate predictions of $s$. Since gradient-boosted collections of trees overfit with large $Ntrees$, we select this hyperparameter via a grid search. To further reduce overfitting, the size of the update at each step is reduced by adding a penalty $0 < \lambda \leq 1$, and predictions are updated with the rule $\hat{s}_{T{}_k}+\lambda\hat{e}_T{}_k = \hat{s}_T{}_{k+1}$. The penalty $\lambda$ is also selected via a grid search.\footnote{The maximum size of each tree in the gradient-boosting algorithm is significantly smaller than in the case of random forests. Consequently, the individual trees in such an ensemble are called \textit{weak learners} (\citet{freund_boosting_1995}, \citet{freund_short_1999}).}

As is customary, the algorithms are trained on the training set, which here contains 80\% of the sample. Each algorithm’s performance is then estimated on the test set, which contains the remaining 20\% of observations. Optimal hyperparameters are chosen via 4-fold cross validation\footnote{Cross-validation is used to mimic the predictive performance of a machine learning model on unseen data. The training set is split into 4 sub-samples. At each step one of the sub-samples is held out while the algorithm is fitted on the remaining 3 sub-samples. Performance is then evaluated on the hold-out sample. Finally, the parameters that are associated with the best predictive performance are chosen.}  on the training set using grid search. Optimal hyperparameters for all the datasets can be found in Appendix Table \ref{tabA1}. Each of these algorithms are implemented using the scikit-learn library in Python (\citet{pedregosa_scikit-learn_2011}). To evaluate the stability of our results across time, where feasible, we train each algorithm on each survey-wave combination separately. 

\subsection{Explanatory variables}

We evaluate each algorithm’s performance for two different sets of explanatory variables. 

As noted above, we first consider a restricted set of variables that are observed in all three of the datasets, which cover basic demographics as well as economic and health variables. We specifically include: sex, age, age-squared, ethnicity, religiosity, number of household members, number of children in the household, marital status, log household income (equivalised used the modified OECD scale), general health status, disability status, body mass index, labour-force status, working hours, home ownership, area of residence, and interview month. A more detailed description of these variables is provided in Appendix Table \ref{tabA3}. These variables are typical in the conventional literature on subjective wellbeing. This restricted set of variables will then allow us to assess the performance of ML algorithms relative to OLS in a standard estimation setting. 

We also evaluate each algorithm on much larger extended sets of explanatory variables. Here, we only use the 2013 Wave of Gallup and SOEP, and Wave 3 of the UKHLS (which covers 2011-2012).\footnote{These waves/years were chosen as they include personality traits in the SOEP and UKHLS.}  Our dataset includes all of the available variables, apart from direct measures of subjective wellbeing (such as domain satisfaction, happiness, or subjective health) or mental health. We also exclude variables with more than 50\% missing values. The resulting Gallup dataset contains 67 variables, and around 450 variables are retained in the SOEP and UKHLS. Missing values for continuous variables are assigned the observed means, while missing values for categorical variables are assigned a new category.\footnote{Processing categorical variables and removing perfectly collinear variables respectively yields 210, 542, and 957 effective explanatory variables in the Gallup, SOEP and UKHLS datasets. } We convert categorical variables into a set of dummies, one for each category.  The full list of variables in this extended set appears in the supplementary material. 

The large number of variables in the extended set produces significant computational burden. At the same time, it is evident that some portion of these variables will have no predictive power for wellbeing. We therefore use LASSO as a device to select the explanatory variables (\citet{tibshirani_regression_1996}, \citet{ahrens_lassopack_2020}).\footnote{Using LASSO on the restricted set of variables produced a similar performance to OLS, with optimal $\lambda=0$.}  We have carried out the estimations on both the full extended set and the post-LASSO extended set. Typically, both approaches perform similarly. For simplicity, we only show results for the approach that performed better in each individual case.

\subsection{Assessing Variable importance}

To answer our second research question, we need to assess how important each explanatory variable is in enabling our algorithms to predict wellbeing. We do so in two ways.

We first use \textit{permutation importances} (PIs) to measure the degree to which each algorithm relies on a given variable in making its predictions (\citet{molnar_interpretable_2022}).\footnote{Shapley values are an alternative option to assess feature importances. However, despite recent advances in reducing the computational complexity of obtaining Shapley values (\citet{Lundberg_machine_2018}), the size of our datasets and models makes the use of Shapley values computationally infeasible (see e.g. \citet{yang_fast_2021}).}  PIs are calculated by randomly shuffling a given variable’s observed values across individuals in the test data and evaluating the extent to which the predictive performance (in terms of R-squared) of a given algorithm falls when permuting the variable’s values. This operation is carried out 10 times. The reported PI is the average change in the R-squared across these 10 iterations. The greater the average fall in the R-squared, the more important is the variable. 

To understand the direction of our variables’ effects we also report \textit{pseudo partial effects} (PPEs). These are calculated by taking the difference in predicted wellbeing after setting each explanatory variable to a given set of values. Specifically, for continuous and ordinal variables we set the variable to the third and first quartile of their distributions and calculate the mean difference in predicted wellbeing. For binary variables (including dummies for all of the categorical variables), we predict wellbeing when setting each individual’s value to either 0 or 1. 

A key advantage of PIs and PPEs is that they can be used with any kind of algorithm, allowing us to compare the way in which each algorithm makes use of the available data.  


\section{Results}
\subsection{Model performance}

We begin with RQ1, \textit{i.e.} whether ML algorithms significantly outperform OLS in predicting wellbeing. As noted, OLS is the standard approach followed in the conventional literature. 

\subsubsection{The Restricted Set of explanatory variables}

\begin{figure}[t]
\centering
\caption{R-squared figures from OLS, GB and RF using the restricted set of variables. The R-squareds are computed using the unseen testing data.}
\label{fig2}
\includegraphics[width=0.9\textwidth]{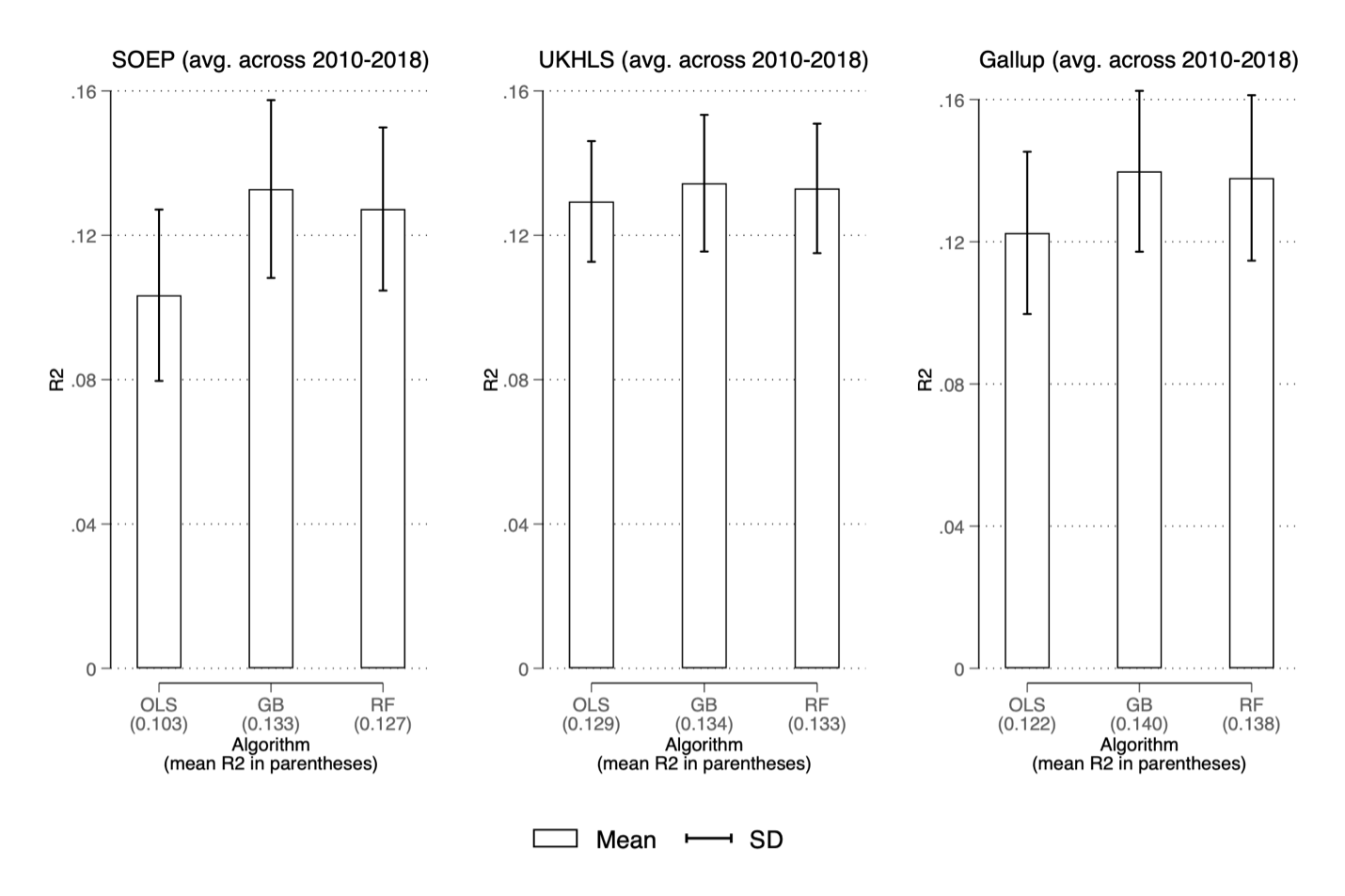}
\end{figure}

We start with the analysis based on the restricted set of covariates, which includes the variables that are typical in many conventional wellbeing estimations. Figure \ref{fig2} depicts the performance of each algorithm on the test-set portion of each dataset. We use R-squared as our primary evaluation metric in order to facilitate the comparison with previous analyses. 

In Figure \ref{fig2} each algorithm is trained separately for each year between 2010 and 2018. The values refer to the average R-squared across these years and their standard deviations. The R-squareds are very similar across datasets, ranging from 0.10 (SOEP) to 0.14 (Gallup). Gradient boosting (GB) and random forests (RF) yield larger R-squared values than OLS in each case. Specifically, random forests yield absolute increases in R-squared of 0.024 (SOEP), 0.004 (UKHLS) and 0.016 (Gallup); the respective improvements from using gradient boosting are slightly larger, with respective R-squared gains of 0.030, 0.005, and 0.018.\footnote{These gains are calculated from the test set, which was not used for training the algorithm. In the training set, \textit{i.e.} the data that is observed by each algorithm, the improvement from performance of the ML algorithms over OLS is larger (see Appendix Figure \ref{fig_a1}). The predictive capacity of the ML algorithms applied to the test set does not then seem to be constrained by underfitting. Of course, performance in the training set is not \textit{per se} indicative of the quality of an algorithm. A decision tree with as many leaves as training individuals would yield an MSE of 0. However, this model would perform extremely poorly when used to assess unseen test data.}  ML algorithms thus do outperform linear regressions, and gradient boosting always outperforms random forests.

\begin{table}[t]
\caption{An illustration of the size of the improvements from using ML.}
\label{tab1}
\begin{tblr}{width=1\textwidth, colspec={lX[c]X[c]X[c]X[c]}}
\hline
       & \begin{tabular}[c]{@{}c@{}}OLS,\\ full\end{tabular} & \begin{tabular}[c]{@{}c@{}}OLS,\\ no health\end{tabular} &             GB & \begin{tabular}[c]{@{}c@{}}GB gain as \% of loss \\ from removing health\end{tabular} \\ \hline
\multicolumn{5}{c}{\textbf{Panel A: Restricted set of variables}}                                                                                                                           \\
SOEP   & 0.103     & 0.075 ($\Delta$=0.028)                                                    & 0.133 ($\Delta$=0.030)                                                 & 107\%                                      \\
UKHLS  & 0.129     & 0.095 ($\Delta$=0.034)                                                     & 0.134 ($\Delta$=0.005)                                                  & 15\%                                       \\
Gallup & 0.122     & 0.093 ($\Delta$=0.029)                                                   & 0.140 ($\Delta$=0.018)                                                 & 62\%                                       \\ \hline
\multicolumn{5}{c}{\textbf{Panel B: Extended set of variables}}                                                                                                                             \\
SOEP   & 0.284     & 0.240 ($\Delta$=0.043)                                                    & 0.318 ($\Delta$=0.035)                                                 & 81\%                                       \\
UKHLS  & 0.215     & 0.197 ($\Delta$=0.018)                                                    & 0.243 ($\Delta$=0.028)                                                 & 155\%                                      \\
Gallup & 0.270     & 0.240 ($\Delta$=0.031)                                                    & 0.280 ($\Delta$=0.018)                                                & 58\%                                       \\ \hline                                    
\end{tblr}
\medskip {\small 
\textbf{Notes:} The figures refer to the R-squared values from the test-set. }
\end{table}

These gain figures considered on their own are hard to interpret. To illustrate the substantive size of these improvements, we compare them to the change in predictive performance when omitting information on respondent’s health status – a key wellbeing predictor – in our baseline OLS regressions. Panel A of Table \ref{tab1} lists the changes in the test-set R-squared of the OLS regression when omitting this information and compares this figure to the gain from using gradient boosting. As benchmarked against the gain from adding health information, the prediction-improvement figure from gradient boosting (as our best ML algorithm) lies between 15\% and 107\%. When evaluated in this way, the gains from using ML do look substantial.

\subsubsection{The Extended Set of explanatory variables}

Adding further explanatory variables should increase our ability to predict wellbeing. Given the greater flexibility of the ML algorithms, we should expect these to benefit more from additional variables than OLS. To test this, we estimate all of our models on the extended sets of variables. As explained in Section 2.3, these extended sets include all of the variables available in the 2013 waves of the SOEP and Gallup, and Wave 3 of the UKHLS. 

Figure \ref{fig3} depicts our main results.\footnote{The results for the training set can be found in Appendix Figure \ref{fig_a2})} The R-squared figure approximately doubles using the extended set for all algorithms, including OLS. The OLS R-squared is now 0.28 for the SOEP, 0.21 in the UKHLS and 0.27 for Gallup. As such, standard economic specifications do not fully exploit the predictive information available in typical large-scale survey data.\footnote{All of these R-squared estimates are obtained using the test set. Hence, these improvements cannot be attributed to a mechanical increase in the share of explained variance due to adding more variables to the model.}

\begin{figure}[t]
\centering
\caption{R-squared figures from OLS, LASSO, GB and RF using the extended set of variables. The R-squareds are computed using the unseen testing data.}
\label{fig3}
\includegraphics[width=0.9\textwidth]{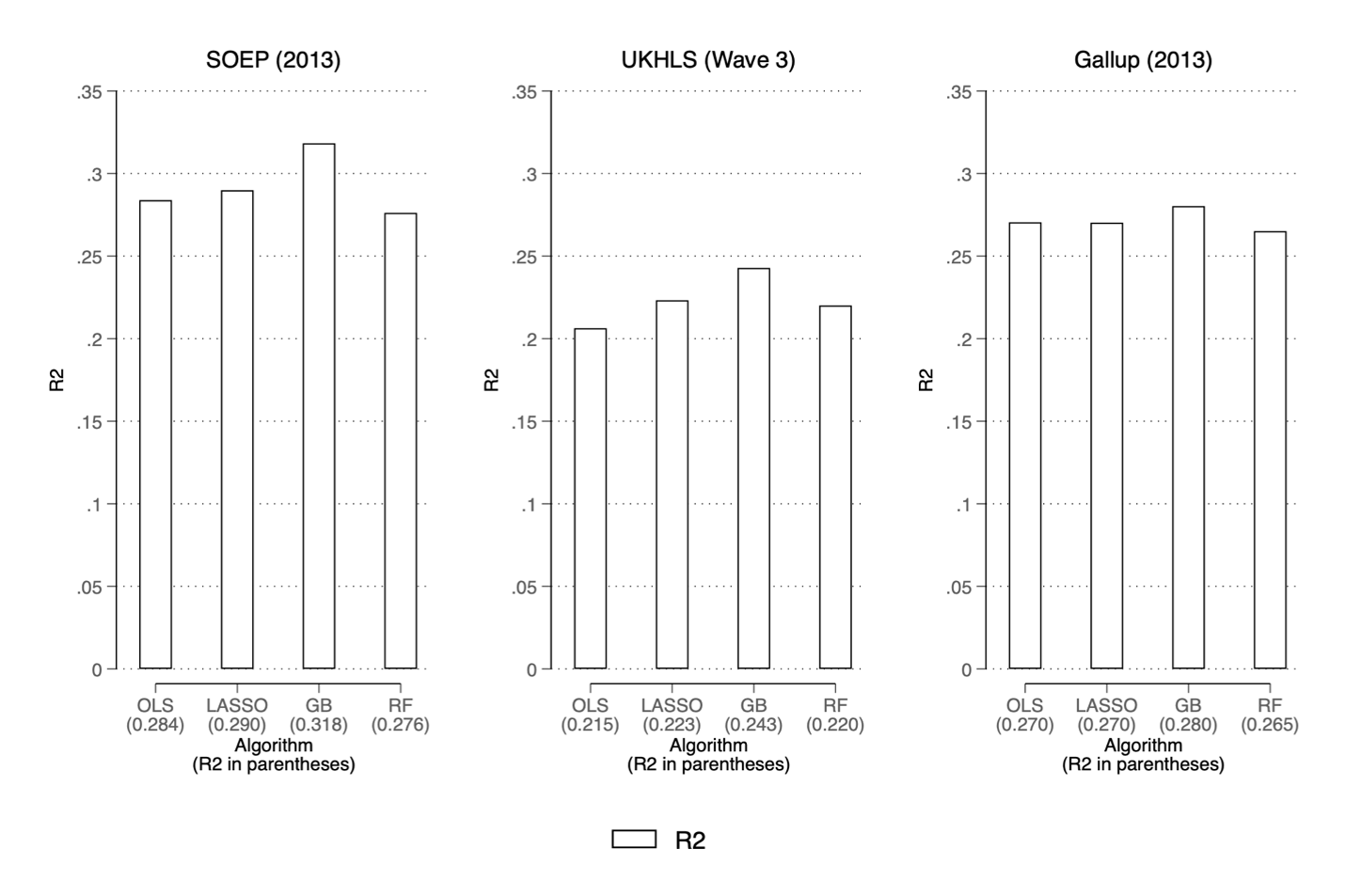}
\end{figure}

Gradient boosting remains the best-performing algorithm and clearly predicts better than OLS. The absolute gain in the R-squared from gradient boosting over OLS is now 0.034, 0.028 and 0.010 for the SOEP, UKHLS and Gallup respectively. Random forests now tend to perform poorly, underperforming OLS for SOEP and Gallup. This has also been observed in other empirical applications where covariates were measured with error (\citet{Reis_2018}).

We again interpret the size of the gains from gradient boosting by comparing them to those from the inclusion of respondents’ health information when using OLS.\footnote{In these extended specifications, there are multiple variables related to health in each dataset. We remove 21, 19 and 12 health-related variables in the Gallup, the SOEP and UKHLS respectively.}  The results in Panel B of Table \ref{tab1} illustrate that these gains are again substantial, being approximately equivalent to the role of health in predicting wellbeing. 

We thus conclude that tree-based ML algorithms can provide improvements in predictive performance over conventional methods. These gains are moderate in absolute terms, but are meaningful when compared to the predictive power of health. However, we also note that these gains are obtained with algorithms that take up to 100 times more time to estimate.\footnote{This figure is based on a comparison between OLS and RF on the Gallup data with the extended dataset.} The use of ML algorithms thus involves a trade-off between computational burden and predictive performance. 

There are multiple reasons that can explain why nonlinear ML methods do not yield a substantial improvement in predicting human wellbeing compared to linear regression. First, most independent variables in the datasets we have used are binary or categorical. Such datasets cannot exhibit non-linearity except by interaction terms between variables. Therefore, if a large number of the variables present only take binary values the ways that improvements can occur with non-linear models are limited. It is possible that non-linear relationships do exist but the variables concerned have a small contribution to the outcome. This is particularly likely if there are many variables contributing to the outcome, as is the case in our extended set of independent variables. Additionally, it may be that the non-linearity is present only at the extremes of the distribution where only few points exist. 

As well as improvements in performance, ML may also indicate new, and potentially-overlooked, variables that are key in explaining subjective wellbeing. The next section explores this idea.

\subsection{Variable importance}

In this section we ask whether the variables that ML identifies as important in predicting life satisfaction correspond to those emphasised in the conventional literature. We do so by estimating variable importances, as discussed in Section 2.4. Our ML-based findings turn out to fit well with the results in previous analyses.

We start by estimating variable importances in the extended dataset, which provides more possibilities for the identification of important variables that do not appear in conventional wellbeing models. Figure \ref{tab2} lists the five most-important variables identified in OLS and GB, which is the best-performing ML algorithm, in each dataset.\footnote{We present the Top-10 most-important variables for OLS, RF and GB in the three datasets in Appendix Table \ref{tabA2}.}  The bars and numerical values refer to permutation importance, \textit{i.e.} the drop in the model’s R-squared when the values of the variable are randomly permuted across respondents. The variables that are negatively associated with average wellbeing are in red, and those with a positive association in green. In all three countries, individual health and interpersonal relationships are among the most-important predictors. As expected, respondents whose health limits their activities are on average less satisfied, while people with fulfilling relationships are typically more satisfied with their lives. The directions of the estimated effects are in line with those in the previous conventional work. ML algorithms and OLS thus generally agree on the direction and approximate size of the most-important variables (see Appendix Table \ref{tabA2} for the effect-size estimates). 

\begin{table}[t]
\captionof{figure}{Permutation importance and pseudo partial effects of OLS and GB on the extended set of variables, 5 most-important variables.}
\label{tab2}
\begin{tabular}{cc}

\multicolumn{2}{c}{\textbf{Panel A: SOEP}}     \\
{ OLS}            & { Gradient Boosting} \\
\begin{minipage}{.5\textwidth}
      \includegraphics[width=\textwidth]{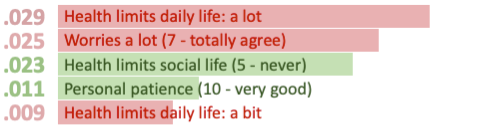}
    \end{minipage}
 & \begin{minipage}{.5\textwidth}
      \includegraphics[width=\textwidth]{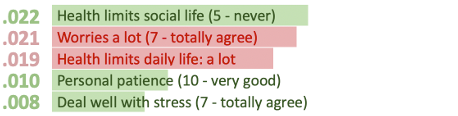}
    \end{minipage}    \\ 
\multicolumn{2}{c}{\textbf{Panel B: UKHLS}}    \\
{ OLS}            & { Gradient Boosting} \\
\begin{minipage}{.5\textwidth}
      \includegraphics[width=\textwidth]{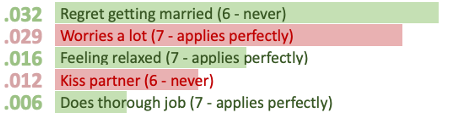}
    \end{minipage}
 & \begin{minipage}{.5\textwidth}
      \includegraphics[width=\textwidth]{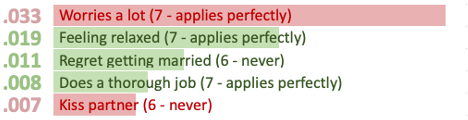}
    \end{minipage}    \\  
\multicolumn{2}{c}{\textbf{Panel C: Gallup}}   \\
{ OLS}            & { Gradient Boosting} \\
\begin{minipage}{.5\textwidth}
      \includegraphics[width=\textwidth]{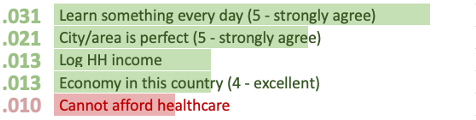}
    \end{minipage}
 & \begin{minipage}{.5\textwidth}
      \includegraphics[width=\textwidth]{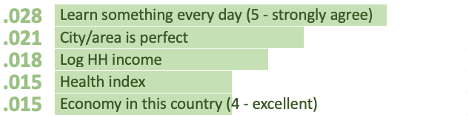}
    \end{minipage}    \\ 
\end{tabular}

\medskip {\small 
\textbf{Notes:} The bars and numerical values represent permutation importance and are coloured red for variables with negative pseudo partial effects and green otherwise. For Likert scale variables, the highest category is reported.}
\end{table}

As a more systematic measure of the degree of agreement between ML and OLS, we calculated the correlations of the ranks (in terms of their permutation importance) of each variable across algorithms and datasets. The results appear in Table \ref{tab3}. There is strong agreement between GB and RF in all three datasets, with the rank correlation figure never falling below 0.79. The correlations with the OLS ranking are somewhat lower, with a minimum value of 0.58 (OLS \textit{vs.} RF in SOEP). Nevertheless, we can strongly reject ($p<0.001$) the null hypothesis that the rankings are uncorrelated, supporting our conclusion that the OLS and ML algorithms are in broad agreement.

\begin{table}[t]
\caption{Correlations between the Permutation Importance ranks in different algorithms.}
\label{tab3}
\begin{tblr}{width=1\textwidth, colspec={lX[c]X[c]X[c]}}
\hline
&  OLS \textit{vs.} GB	& OLS \textit{vs.} RF	& GB \textit{vs.} RF \hline\\ 
SOEP &	0.70 &	0.58 &	0.79 \\
UKHLS &	0.75 &	0.67 &	0.86 \\
Gallup &	0.86 &	0.69 &	0.82 \\
 \hline                                     
\end{tblr}

\medskip {\small 
\textbf{Notes:} The correlation figures refer to the Top-100 variables (using the OLS ranking). These are Spearman rank correlations. }
\end{table}

Apart from the conventional variables used in wellbeing analysis, such as health and interpersonal relationships, the algorithms also identify personality traits as important predictors in the UKHLS and SOEP. Personality traits, unfortunately, do not appear in the Gallup survey. In the UK data, measures associated with (the absence of) neuroticism (\textit{i.e.} worrying, and feeling relaxed) appear in the Top-3. In German data, worrying a lot, being able to deal with stress, and patience are among the most-important variables in all empirical approaches. This is line with previous research underlining the potential advantages of including personality traits in wellbeing regressions (\citet{ferrericarbonell_how_2004}, \citet{proto_covid-19_2021}).

Beyond these similarities, there are some cross-country differences. The most striking refer to the importance of financial factors. These are important in the US (e.g., HH income and being able to pay for healthcare) but not in the other countries. To see whether this is a genuine finding or a consequence of differences in variable availability across countries, we carry out the same analysis using the restricted set (for which we have a common set of variables). When we do so, the cross-country differences in the importance of income largely disappear. As shown in Appendix Table \ref{tabA4}, the most-important variables identified in these harmonised datasets are very similar across the three countries. They include health, income, marital and employment status, as well as home-ownership – which is a proxy for wealth – and age. Sex and ethnicity are only important in the US. Education is among the most important factors in the US and Germany, but not in the UK.

\subsection{Additional analyses and robustness tests}
\subsubsection{Wellbeing by age and income}

The preceding section concluded that the kinds of variables that machine learning finds to be important - and the estimated direction of their association with wellbeing - are largely in line with the results in the conventional literature. We here present a detailed analysis of two variables that have attracted a great deal of interest in the conventional literature: age and income. In OLS estimation, the functional forms associated with these two variables are imposed by the analyst, while they are instead freely estimated in our tree-based ML algorithms. 

The results appear in Figure \ref{tab4} and Appendix Figure \ref{fig_a3}. In the OLS estimation, illustrated in blue, we assume a quadratic form for age, and a log-linear functional form for income, which are very common functional forms in this literature. The relationships for RF are in red, and those for GB in green.

\begin{table}[]
\captionof{figure}{The mean effects of age and household income on wellbeing, restricted set of variables.}
\label{tab4}
\begin{tabular}{cc}
\multicolumn{2}{c}{\textbf{Panel A: SOEP}}     \\
\begin{minipage}{.5\textwidth}
      \includegraphics[width=\textwidth]{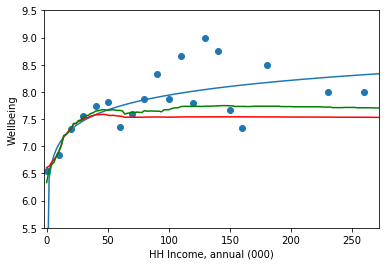}
    \end{minipage}
 & \begin{minipage}{.5\textwidth}
      \includegraphics[width=\textwidth]{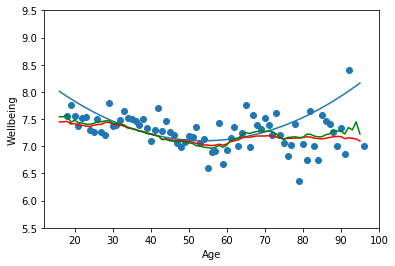}
    \end{minipage}    \\ 
\multicolumn{2}{c}{\textbf{Panel B: UKHLS}}    \\
\begin{minipage}{.5\textwidth}
      \includegraphics[width=\textwidth]{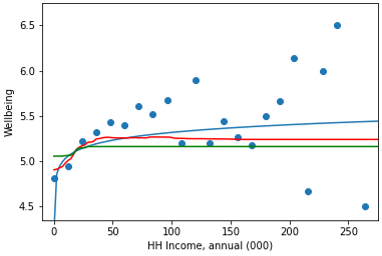}
    \end{minipage}
 & \begin{minipage}{.5\textwidth}
      \includegraphics[width=\textwidth]{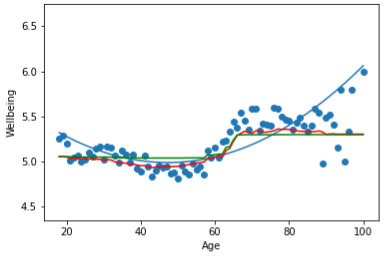}
    \end{minipage}    \\  
\multicolumn{2}{c}{\textbf{Panel C: Gallup}}   \\
\begin{minipage}{.5\textwidth}
      \includegraphics[width=\textwidth]{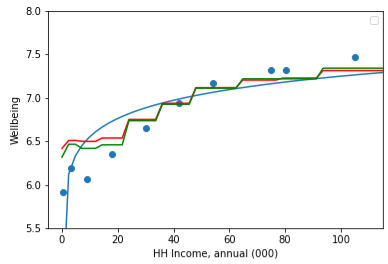}
    \end{minipage}
 & \begin{minipage}{.5\textwidth}
      \includegraphics[width=\textwidth]{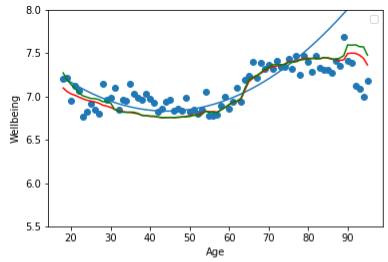}
    \end{minipage}    \\  
\multicolumn{2}{c}{\begin{minipage}{.7\textwidth}
      \includegraphics[width=\textwidth]{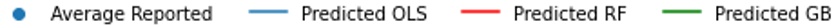}
    \end{minipage}}   \\    
\end{tabular}

\medskip {\small 
\textbf{Notes:} For the UKHLS and the SOEP annual income is constrained to be less than or equal to a figure of 250 000 in the local currency. This covers over 99.9\% of the income distribution in both countries. In SOEP and UKHLS, incomes are recorded as a continuous variable and equivalence-scale adjusted HH income is used for the analysis. Income data in Gallup is collected in income bands, and household size data was not collected in 2013. We here thus use non-adjusted HH income data. }
\end{table}

For low to medium incomes, both ML algorithms track the assumed log-normal functional form remarkably closely, in line with the conventional literature. However, once we reach relatively-high equivalised annual income figures, above 50,000 EUR in the SOEP or 40,000 GBP in the UKHLS, the ML algorithms suggest that wellbeing no longer increases with income. We cannot confirm this finding in the US, as income in Gallup appears in bands with the highest band being 100,000 USD or above. In 2013, 100,000 USD were approximately equivalent to 70,000 GBP or 78,000 EUR.\footnote{\url{https://data.oecd.org/conversion/purchasing-power-parities-ppp.htm }}  In addition, the Gallup 2013 wave did not collect data on household size. As a result, household income in Gallup is not directly comparable to the adjusted equivalent incomes in SOEP and UKHLS. Given these caveats, we do not find evidence of satiation in the US data. Our ML findings are therefore in line with previous work on wellbeing using US data (\citet{kahneman_high_2010}, \citet{killingsworth_experienced_2021}). 

With respect to the relationship between age and wellbeing, our ML estimations replicate the well-known approximate U-shape up to age 70 (e.g. \citet{cheng_longitudinal_2017}), which is more pronounced in the US. However, unlike the smooth U-shape assumed in the OLS approach, we find a much more pronounced “kink” at around age 65 for each dataset and ML-algorithm. We suspect that this kink reflects the gains in wellbeing following on from retirement (\citet{gorry_does_2018}, \citet{wetzel_transition_2016}).

\subsubsection{Positive and negative affect}

We also evaluate the performance of gradient boosting and random forests on measures of positive and negative affect. The results show that our findings are not specific to the use of life evaluations as the measure of subjective wellbeing, but generalise to affect (or mood).  
In the 2013 Gallup data, positive affect is measured by the average figure from dummy variables indicating whether the respondent felt happiness or joy, or smiled during the previous day. Negative affect is calculated analogously from dummies indicating pain, worry, sadness, and anger. In the German SOEP, positive affect is the self-reported frequency of being happy in the last 4 weeks (on a 1 to 5 scale), and negative affect as the average of the self-reported frequency from three questions about being angry, sad, or worried in the last four weeks (all measured on a 1 to 5 scale). The UKHLS dataset does not contain comparable affect data and is not used in this part of analysis. 

The results for the Gallup data appear in the top panels of Figure \ref{fig_a4} and Table \ref{tabA5}. It is notable that negative affect is easier to predict than positive affect. This finding holds across algorithms, with R-squared figures ranging from 0.423 and 0.464 for negative affect, and between 0.261 and 0.296 for positive affect. Random forests and gradient boosting outperform both OLS and LASSO. As was the case for life evaluations, gradient boosting again performs the best, with gains in R-squared over OLS of 0.041 for negative affect and 0.036 for positive affect. Regarding variable importances, Table \ref{tabA5} shows that good health is even more important for predicting positive and negative affect in the Gallup data than it was for life evaluation. Moreover, in line with previous work (e.g. \citet{kahneman_high_2010}), variables relating to material conditions – like income – do not feature in the list of the set of most-important variables when modelling affect. 

As shown in Table \ref{tabA6} and the bottom panels of Figure \ref{fig_a4}, the results are qualitatively similar in the German data: gradient boosting again performs best, and positive affect is harder to predict than negative affect.  

\subsubsection{Panel data}

Our main findings regarding the ML estimation of wellbeing are also robust to exploiting the panel dimension of the German SOEP and the UKHLS. As there is no standard procedure for the introduction of individual fixed effects in the ML algorithms that we use, we implement an approach similar to the Mundlak correction for linear models (\citet{mundlak_pooling_1978}, \citet{wooldridge_econometric_2010}): we pool all years of the UKHLS and SOEP data, demean all covariates at the individual level and include both an individual’s average value over time of each covariate as well as their year-specific deviations from their individual mean. The level of wellbeing is the dependent variable, as was the case in the analysis above. 

The relative predictive performance of the OLS and ML in the pooled dataset is similar to the findings for individual years. In the UKHLS, the OLS R-squared is 0.140. The use of RF produces a small improvement, with the R-squared increasing to 0.143. Gradient boosting provides a further improvement, yielding an R-squared of 0.150. In the German SOEP, the OLS R-Squared is 0.12, with once again both the random forest and gradient boosting leading to better R-Squared figures of, respectively, 0.15 and 0.16. As shown in Tables \ref{tabA7} and \ref{tabA8}, the most important variables predicting the level of wellbeing are almost exclusively the average values of the individual covariates. One exception in both the UKHLS and SOEP is the \textit{Health limits activities} variable. As such, deviations in individual health status (from their average value) seem to be important for the level of individual wellbeing.

\subsection{Discussion}

We draw three main conclusions from our analysis above.

First, tree-based ML approaches do indeed perform better at predicting wellbeing than more-conventional linear models. Although the gains in R-squared we obtain are modest in absolute terms, they are comparable with – and sometimes exceed – the extent to which information on respondents’ health can improve wellbeing predictions. Comparing the algorithms we consider, gradient boosting consistently outperforms random forests. 

Second, when we use all of the non-wellbeing variables that are available in each dataset as predictors, we more than double the explained variation in wellbeing for all of the estimation procedures that we analyse. This extended set of variables produces R-squared figures of around 0.3. These values look to be the maximum achievable with the current survey data. 

Third, almost all of the variables that turn out to be important in the specifications using of all the available data relate to health, economic conditions, personality traits, and personal relationships. This purely data-driven process thus picks out the same core determinants of wellbeing as have been identified in the conventional literature. In that sense, machine-learning approaches validate the previous human-guided search for the determinants of wellbeing. This looks to be good news for the field. 

We see two directions for future research.

The first is to further explore the capabilities of ML models. We have focused our analysis here on tree-based methods, which are powerful algorithms that perform well in multiple contexts. However, given the specificities of wellbeing data, we might find further improvements by using other algorithms (e.g. Kernel Ridge,  \citet{Vovk2013}), or by using a combination of theory-based modelling and algorithmic approaches.
Another potential approach is using a combination of unsupervised and supervised learning. For example, it might be possible to separate the whole dataset into overapping clusters of individuals chosen based on subsets of independent variables. Then, the predictive performance of non-linear ML models could be substantially higher when applied to such clusters, as compared to using one global model for the whole dataset as done in our work. Moreover, we have currently only focused on identifying variables that are key for the successful prediction of wellbeing. A natural next step is to extend the use of ML-based algorithms to investigate the variables that are most important for wellbeing in a causal sense (\citet{wager_estimation_2018}). 

Second, our analyses focused on rich Western countries. As such, it remains an open question whether our findings would also hold in a more global setting, e.g. in countries where material needs are much more acute. Insofar as there may be greater scope for improving wellbeing in low- and middle-income countries (\citet{helliwell_happiness_2022}), applying ML approaches in this setting may be particularly valuable going forward.  

\newpage

\printbibliography[heading=bibintoc]

\newpage
\appendix
\section*{Appendix}
\setcounter{figure}{0}
\setcounter{table}{0}
\counterwithin{figure}{section}
\counterwithin{table}{section}
\renewcommand\thefigure{A\arabic{figure}}
\renewcommand\thetable{A\arabic{table}}

\begin{table}[H]
\caption{Optimal hyperparameters used in the extended specifications (post-LASSO extended specification in parentheses)}
\label{tabA1}
\begin{tblr}{width=1\textwidth, colspec={lX[c]X[c]X[c]}}
\hline
                  & \multicolumn{3}{c}{\textbf{Panel A: Random Forest}}     \\
                  & SOEP          & Gallup           & UKHLS       \\ \hline
MaxDepth          & 96 (70)       & 70 (70)          & 30 (20)     \\
Nvars             & 225 (65)      & 80 (80)          & 400 (130)   \\
Ntrees            & 1000 (1000)   & 1000 (1000)      & 1000 (1000) \\
MinLeaf           & 1 (1)         & 5 (5)            & 15 (5)      \\ \hline
                  & \multicolumn{3}{c}{\textbf{Panel B: Gradient Boosting}} \\
                  & SOEP          & Gallup           & UKHLS       \\ \hline
MaxDepth          & 8 (8)         & 3 (3)            & 5 (7)       \\
Nvars             & 75 (30)       & 40 (40)          & 100 (30)    \\
Ntrees            & 6000 (2000)   & 16000 (16000)    & 2000 (2000) \\
MinLeaf           & 1 (1)         & 1 (1)            & 1 (1)       \\
Learning rate ($\lambda$) & 0.005 (0.01)  & 0.0063 (0.0063)  & 0.01 (0.01) \\ \hline
\end{tblr}

\vspace{0.1cm}

\linespread{0.9}\small \textbf{Notes:} 
Hyperparameters are identified via a grid search by minimizing the average MSE across 4 folds of cross-validation. \textit{MaxDepth} is the maximum depth of each branch of each tree. \textit{Nvars} is the maximum number of randomly-picked variables used to perform splits within each tree. \textit{MinLeaf} is the minimum number of training individuals that must be in each leaf of a given tree (fixed to 1 for gradient boosting). \textit{Ntrees} is the number of trees fitted (fixed to 1,000 for random forests). The learning rate ($\lambda$) is the rate at which predictions are updated (only applicable to gradient boosting).
\end{table}

\begin{table}[H]
\caption{List of variables in the restricted set.}
\label{tabA3}
\begin{tblr}{width=1\textwidth, colspec={X[1.5,l]X[r]X[r]X[r]}}
\hline
Variable                       & SOEP                                                                                & UKHLS                                        & Gallup                                 \\ \hline
Age                            & 16 - 105                                                                            & 18 - 103                                     & 18 - 99                                \\
Area of residence              & 16 distinct values                                                                  & 12 regions                                   & 51 distinct values                     \\
BMI                            & 11.10 - 84.50                                                                       & 11.80 - 74.20                                & 7.19 - 152.56                          \\
Disability status              & Binary                                                                              & Binary                                       & n.a.                                   \\
Education                      & 18 - 7  (years of education)                                                        & 6 distinct values                            & 6 distinct values                      \\
Labour-force status              & Binary                                                                              & 12 distinct values                           & 4 distinct values                      \\
Equivalised Log HH income      & 0 - 13.88                                                                           & -0.80 - 12.52                                & 3.40 - 9.90                            \\
Ethnicity/Migration background & 3 distinct values (migration background) & 18 distinct values (ethnicity)               & 5 distinct values (ethnicity)          \\
Health                         & 0 – 396 (doctor visits in prev. year)                                               & Health limits activities (3 distinct values) & Binary (self-assessed health problems) \\
Housing status         & 4 distinct values                                                                   & 6 distinct values                            & n.a.                                   \\
Marital status                 & 5 distinct values                                                                   & 10 distinct values                           & 6 distinct values                      \\
Month of interview             & 12 distinct values                                                                  & 24 distinct values                           & 12 distinct values                     \\
Number of children in HH       & 0 - 11                                                                              & 0 - 9                                        & 0 - 15                                 \\
Number of people in HH         & 1 - 16                                                                              & 1 - 16                                       & 1 - 99                                 \\
Religion                       & 10 distinct values                                                                  & Binary                                       & 8 distinct values                      \\
Sex                            & Binary                                                                              & Binary                                       & Binary                                 \\
Working hours                  & 0 - 6669                                                                            & 0 - 180                                      & 4 distinct values                      \\ \hline
\end{tblr}
\end{table}
\vspace{-0.4cm}
\linespread{0.9}
\small\textbf{Notes:} For continuous variables, the range is reported. For \underline{SOEP}, possible values for the categorical variables are: \textit{Area of residence}: Each of the 16 Bundesländer. \textit{Ethnicity/Migration background:} No migration background, Direct migration background, Indirect migration background. \textit{Housing status:} Main Tenant, Sub-Tenant, Owner, Nursing Home/ Retirement Community. \textit{Marital status:} Married, Single, Widowed, Separated, Divorced. \textit{Religion:} Catholic, Protestant, Christian Orthodox, Other Christian, Muslim, Muslim (Shiite), Muslim (Sunnite), Muslim (Alevite), Other, No religion. For \underline{UKHLS}, possible values for the categorical variables are: \textit{Area of residence:} North East, North West, Yorkshire and the Humber, East Midlands, West Midlands, East of England, London, South East, South West, Wales, Scotland, Northern Ireland. \textit{Education:} Degree, Other higher degree, A-level etc, GCSE etc., Other qualification, No qualification. \textit{Labour-force status:} Self-employed, Paid employment(ft/pt), Unemployed, Retired, On maternity leave, Family care or home, Full-time student, LT sick or disabled, Govt training scheme, Unpaid, family business, On apprenticeship, Doing something else.\textit{ Ethnicity:} British/English/Scottish/Welsh/Northern irish, Irish, Gypsy or Irish traveller, Any other white background, White and black caribbean, White and black african, White and asian, Any other mixed background, Indian, Pakistani, Bangladeshi, Chinese, Any other asian background, Caribbean, African, Any other black background, Arab, Any other ethnic group. \textit{Health limits moderate activities:} Yes, a lot; Yes, a little; No, not at all. \textit{Housing status:} Owned outright, Owned/being bought on mortgage, Shared ownership (part-owned part-rented), Rented, Rent free, Other. \textit{Marital status:} Single and never married/in civil partnership, Married, In a registered same-sex civil partnership, Separated but legally married, Divorced, Widowed, Separated from civil partner, A former civil partner, A surviving civil partner, Living as couple. For \underline{Gallup}, possible values for the categorical variables are: \textit{Area of residence:} 51 States. \textit{Education:} Less than high school, High school, Technical/Vocational school, Some college, College graduate, Post-graduate. \textit{Labour-force status: }Employed, Self-employed, Employed and self-employed, not employed. \textit{Ethnicity:} White, Other, Black, Asian, Hispanic. \textit{Marital status:} Single, Married, Separated, Divorced, Widowed, Living with partner (not married). \textit{Religion:} Protestant, Catholic, Jewish, Muslim, Mormon, Other Christian, Other, No religion.  \textit{Working hours:} 30 or more hours per week, 15 to 29 hours per week, 5 to 14 hours per week, less than 5 hours per week.  
\linespread{1.3}

\begin{table}[H]
\caption{Permutation Importance (PI) and Pseudo Partial Effects (PPE) in OLS, RF and GB on the Extended Set of variables: the 10 most-important variables.}
\label{tabA2}
\resizebox{1\textwidth}{!}{\begin{tabular}{llrrlrrlrrr}
\hline
\multicolumn{1}{c}{} & \multicolumn{3}{c}{OLS}                                                              & \multicolumn{3}{c}{Random forest}                                                    & \multicolumn{4}{c}{Gradient boosting}                                                                  \\
\multicolumn{1}{c}{} & \multicolumn{1}{c}{Variable name} & \multicolumn{1}{c}{PI} & \multicolumn{1}{c}{PPE} & \multicolumn{1}{c}{Variable name} & \multicolumn{1}{c}{PI} & \multicolumn{1}{c}{PPE} & \multicolumn{2}{c}{Variable name}                   & \multicolumn{1}{c}{PI} & \multicolumn{1}{c}{PPE} \\ \hline
\multicolumn{11}{c}{\textbf{Panel A: SOEP}}                                                                                                                                                                                                                                                                          \\
1                    & Health limits daily life: a lot   & .029                   & -.780                   & Health limits social life         & .032                   & .154                    & \multicolumn{2}{l}{Health limits social life}       & .022                   & .172                    \\
2                    & Worry a lot                       & .025                   & -.146                   & Health limits daily life: a lot   & .028                   & -.742                   & \multicolumn{2}{l}{Worry a lot}                     & .021                   & -.100                   \\
3                    & Health limits social life         & .023                   & .187                    & Worry a lot                       & .020                   & -.113                   & \multicolumn{2}{l}{Health limits daily life: a lot} & .019                   & -.628                   \\
4                    & Personal patience                 & .011                   & .129                    & HH income                         & .018                   & .202                    & \multicolumn{2}{l}{Personal patience}               & .010                   & .174                    \\
5                    & Health limits daily life: a bit   & .009                   & -.266                   & Deal well with stress             & .015                   & .160                    & \multicolumn{2}{l}{Deal well with stress}           & .008                   & .128                    \\
6                    & Partner in HH                     & .008                   & .222                    & Personal patience                 & .008                   & .106                    & \multicolumn{2}{l}{Health limits daily life: a bit} & .006                   & -.220                   \\
7                    & No monthly savings                & .008                   & -.186                   & No annual holiday trip            & .007                   & -.114                   & \multicolumn{2}{l}{Partner in HH}                   & .006                   & .152                    \\
8                    & Deal well with stress             & .006                   & .080                    & No monthly savings                & .007                   & -.110                   & \multicolumn{2}{l}{Risk tolerance}                  & .006                   & .036                    \\
9                    & House needs repair                & .005                   & -.126                   & Not unemployed                    & .006                   & .303                    & \multicolumn{2}{l}{HH income}                       & .006                   & .152                    \\
10                   & Hours of sleep on workday         & .004                   & .077                    & Unemployment benefit              & .005                   & -000                    & \multicolumn{2}{l}{Number of doctor visits}         & .006                   & -.086                   \\ \hline
\multicolumn{11}{c}{\textbf{Panel B: UKHLS}}  \\
1                    & Regret getting married            & .032                   & .418                    & Worries a lot (Big 5)             & .030                   & -.146                   & \multicolumn{2}{l}{Worries a lot (Big 5)}           & .033                   & -.188                   \\
2                    & Worries a lot (Big 5)             & .029                   & -.274                   & Feeling relaxed (Big 5)           & .027                   & .238                    & \multicolumn{2}{l}{Feeling relaxed (Big 5)}         & .019                   & .212                    \\
3                    & Feeling relaxed (Big 5)           & .016                   & .240                    & Health limits kind of work        & .009                   & .040                    & \multicolumn{2}{l}{Regret getting married}          & .011                   & .209                    \\
4                    & Kiss partner                      & .012                   & -.218                   & Belong to neighbourhood           & .009                   & -.179                   & \multicolumn{2}{l}{Does a thorough job (Big5)}      & .008                   & .069                    \\
5                    & Does thorough job (Big 5)         & .006                   & .112                    & Age squared                       & .009                   & .007                    & \multicolumn{2}{l}{Kiss partner}                    & .007                   & -.110                   \\
6                    & Share interests w. partner        & .006                   & -.161                   & Regret getting married            & .009                   & .137                    & \multicolumn{2}{l}{Age squared}                     & .007                   & .002                    \\
7                    & Belong to neighbourhood           & .005                   & -.107                   & Health limits work amount         & .008                   & .032                    & \multicolumn{2}{l}{Health limits kind of work}      & .007                   & .053                    \\
8                    & Sociable (Big 5)                  & .005                   & .094                    & Does thorough job (Big 5)         & .007                   & .053                    & \multicolumn{2}{l}{Health limits work amount}       & .006                   & .049                    \\
9                    & Health limits work amount         & .005                   & .070                    & Consider divorce (never)          & .006                   & .106                    & \multicolumn{2}{l}{Belong to neighbourhood}         & .006                   & -.162                   \\
10                   & Long term sick or disabled        & .005                   & -.420                   & Sociable (Big 5)                  & .006                   & .081                    & \multicolumn{2}{l}{Sociable (Big 5)}                & .006                   & .126                    \\ \hline
\multicolumn{11}{c}{\textbf{Panel C: Gallup}}                                                                                                                                                                                                                                                                        \\
1                    & Learn something every day         & .031                   & .43                     & Learn something every day         & .033                   & .34                     & \multicolumn{2}{l}{Learn something every day}       & .028                   & .35                     \\
2                    & City/area is perfect              & .021                   & .32                     & City/area is perfect              & .026                   & .42                     & \multicolumn{2}{l}{City/area is perfect}            & .021                   & .39                     \\
3                    & Log HH income                     & .013                   & .15                     & Log HH income                     & .021                   & .30                     & \multicolumn{2}{l}{Log HH income}                   & .018                   & .26                     \\
4                    & Economy in this country           & .013                   & .21                     & Cannot afford healthcare          & .021                   & -.54                    & \multicolumn{2}{l}{Health index}                    & .015                   & .16                     \\
5                    & Cannot afford healthcare          & .010                   & -.38                    & Economy in this country           & .015                   & .21                     & \multicolumn{2}{l}{Economy in this country}         & .015                   & .22                     \\
6                    & Health limits activities          & .010                   & -.04                    & Physical health index             & .013                   & .15                     & \multicolumn{2}{l}{Cannot afford healthcare}        & .013                   & -.40                    \\
7                    & Health encouragement              & .010                   & .12                     & Health limits activities          & .010                   & -.03                    & \multicolumn{2}{l}{Health encouragement}            & .008                   & .17                     \\
8                    & Physical health index             & .010                   & .14                     & Health encouragement              & .010                   & .17                     & \multicolumn{2}{l}{Health limits activities}        & .008                   & -.01                    \\
9                    & Female                            & .008                   & .24                     & Female                            & .005                   & .13                     & \multicolumn{2}{l}{Age and age-squared}             & .005                   & .03                     \\
10                   & Ever diag. w depression           & .008                   & -.28                    & Ever diag. w. depression          & .005                   & -.16                    & \multicolumn{2}{l}{Female}                          & .005                   & .25                     \\ \hline
\end{tabular}}
\vspace{0.1cm}

\linespread{0.9}\small
 \textbf{Notes:} The following variables are shown. \underline{SOEP:} Dummies: \textit{Health limits daily life a lot}, \textit{Health limits daily life a bit}, \textit{Partner in HH}, \textit{No monthly savings}, \textit{Not unemployed}, \textit{No emergency reserves}, and \textit{No annual holiday trip}. Likert scales: \textit{Limited socially due to health} (1 – always to 5 – never), \textit{Worries a lot} and \textit{Deals well with stress} (1 – not at all to 7 – totally agree), \textit{Personal patience} (0 – very bad to 10 – very good), \textit{House needs repair} (1 – in good condition, 3 – needs major renovation). Continuous: \textit{Log HH income}, \textit{Hours of sleep}, \textit{Number of Doctor visits}, \textit{Risk Tolerance and Unemployment Benefit}. \underline{UKHLS:} Dummies: \textit{Health not limiting activities}. Likert scales: \textit{Pain interferes with work} (1 – not at all to 5 – extremely), \textit{Regret getting married}, \textit{Share interests w. partner}, \textit{Consider divorce} and \textit{Kiss partner} (1 – all the time, 6 – never), \textit{Health limits work amount} and \textit{Health limits kind of work} (1 – all of the time, 5 – none of the time); Big 5 traits, including\textit{ Worries a lot}, \textit{Feeling relaxed}, \textit{Does thorough job}, \textit{Is sociable} (1 – does not apply to 7 – applies perfectly), \textit{Belong to neighbourhood} (1 – strongly agree – 5 strongly disagree). Continuous: \textit{Age squared}. \underline{Gallup:} Dummies: \textit{Cannot afford healthcare}, \textit{Female}, \textit{Ever diagnosed with depression}. Likert scales:\textit{ Learn something every day}, \textit{City/area is perfect} and\textit{ Receives Health encouragement} (1 – strongly disagree, 5 – strongly agree), \textit{Economy in this country} (1 – poor to 4 – Excellent), \textit{Health limits activities in the last month} (0 to 30 days). Continuous: \textit{Age}, \textit{age squared}, \textit{Log HH income}, \textit{Physical health index}.
\end{table}

\newpage
\begin{table}[H]
\caption{Permutation Importance (PI) and Pseudo Partial Effect (PPE) in OLS, RF and GB on the Restricted Set of variables: the 10 most-important variables.}
\label{tabA4}
\resizebox{1\textwidth}{!}{\begin{tabular}{llrrlrrlrr}
\hline
\multicolumn{1}{c}{} & \multicolumn{3}{c}{OLS}                                                              & \multicolumn{3}{c}{Random forest}                                                    & \multicolumn{3}{c}{Gradient boosting}                                                                  \\
\multicolumn{1}{c}{} & \multicolumn{1}{c}{Variable name} & \multicolumn{1}{c}{PI} & \multicolumn{1}{c}{PPE} & \multicolumn{1}{c}{Variable name} & \multicolumn{1}{c}{PI} & \multicolumn{1}{c}{PPE} & \multicolumn{1}{c}{Variable name}                   & \multicolumn{1}{c}{PI} & \multicolumn{1}{c}{PPE} \\ \hline
\multicolumn{10}{c}{\textbf{Panel A: SOEP}}                                                   \\
1  & Age and age-squared        & .10 & -1.70 & Adjusted Income         & .13 & .27  & Adjusted Income            & .14 & .46  \\
2  & Adjusted Income            & .10 & .26   & Age and age-squared     & .12 & -.14 & Age and age-squared        & .13 & -.18 \\
3  & Number of doctor visits    & .08 & -.14  & Number of doctor visits & .11 & -.28 & Number of doctor visits    & .12 & -.63 \\
4  & Marital Status - Single    & .07 & -.40  & Disability Status       & .04 & -.40 & Disability Status          & .03 & -.45 \\
5  & N of children in HH        & .06 & .30   & N of children in HH     & .03 & .07  & Working hours              & .02 & -.29 \\
6  & Disability Status          & .04 & -.52  & N of people in HH       & .03 & .02  & N of years of education    & .02 & .17  \\
7  & N of people in HH          & .03 & -.17  & N of years of education & .02 & .07  & N of children in the HH    & .02 & .08  \\
8  & N of years of education    & .03 & .11   & House Ownership: Owner  & .02 & .12  & N of people in HH          & .02 & -.16 \\
9  & Marital Status – Divorced  & .02 & -.38  & Working hours           & .01 & .04  & Marital Status – Single    & .02 & -.19 \\
10 & Marital Status - Separated & .02 & -.74  & BMI                     & .01 & -.02 & Marital Status - Separated & .01 & -.53 \\ \hline
\multicolumn{10}{c}{\textbf{Panel B: UKHLS}}                                                                                                                                             \\
1  & Health limits activities: a lot & .024 & -.670 & Age                             & .040 & .052  & LT sick or disabled (empl.)     & .018 & -.587 \\
2  & Single                          & .020 & -.336 & HH income                       & .015 & .161  & Age                             & .015 & .052  \\
3  & LT sick or disabled (empl.)     & .017 & -.797 & Health limits activities: a lot & .014 & -.377 & Health limits activities: a lot & .012 & -.377 \\
4  & Age                             & .018 & .015  & Not disabled (health)           & .014 & .215  & Not disabled (health)           & .010 & .215  \\
5  & Health limits activities: a bit & .014 & -.327 & Health limits activities: a bit & .012 & -.226 & Renting house                   & .007 & -.106 \\
6  & Not disabled (health)           & .011 & .240  & LT sick or disabled (empl.)     & .011 & -.587 & Health limits activities: a bit & .007 & -.226 \\
7  & Retired                         & .010 & .235  & Unemployed                      & .006 & -.193 & HH income                       & .006 & .161  \\
8  & Renting house                   & .008 & -.208 & Renting house                   & .005 & -.106 & Unemployed                      & .006 & -.193 \\
9  & Unemployed                      & .008 & -.343 & Single                          & .005 & -.136 & Retired                         & .005 & .099  \\
10 & HH income                       & .008 & .083  & Retired                         & .003 & .099  & Single                          & .003 & -.136 \\\hline
\multicolumn{10}{c}{\textbf{Panel C: Gallup}}                                                                                                                                            \\
1  & Health limits activities & .064 & .84  & HH income                & .062 & .48  & HH income                & .067 & .48  \\
2  & HH income                & .049 & .30  & Health limits activities & .057 & .69  & Health limits activities & .054 & .71  \\
3  & Post-graduate education  & .026 & .58  & Age and age-squared      & .046 & .43  & Age and age-squared      & .041 & .44  \\
4  & Married                  & .013 & .33  & Married                  & .013 & .26  & Married                  & .013 & .27  \\
5  & College Graduate         & .010 & .37  & Female                   & .010 & .23  & Female                   & .013 & .29  \\
6  & Female                   & .010 & .29  & Post-graduate education  & .008 & .43  & Post-graduate education  & .008 & .34  \\
7  & Age and age-squared      & .008 & .24  & Body Mass Index          & .005 & .29  & Body Mass Index          & .005 & -.12 \\
8  & Hispanic                 & .003 & .28  & Working Hours Missing    & .005 & -.12 & Hispanic                 & .003 & .15  \\
9  & Atheist                  & .003 & -.19 & Hispanic                 & .003 & .06  & Black                    & .003 & .10  \\
10 & High school graduate     & .003 & .17  & Asian                    & .003 & .02  & Working Hours Missing    & .003 & -.06 \\ \hline
\end{tabular}}
\vspace{0.1cm}

\linespread{0.9}\small
 \textbf{Notes:} The total set of variables available in the restricted set appears in Table \ref{tabA1}.  
\end{table}

\newpage
\begin{table}[H]
\caption{Permutation Importance (PI) and Pseudo Partial Effect (PPE) in OLS, RF and GB for positive and negative affect: the top 10 most-important variables (using 2013 Gallup data with the Extended Set of variables).}
\label{tabA5}
\resizebox{1\textwidth}{!}{\begin{tabular}{llrrlrrlrr}
\hline
\multicolumn{1}{c}{} & \multicolumn{3}{c}{OLS}                                                              & \multicolumn{3}{c}{Random forest}                                                    & \multicolumn{3}{c}{Gradient boosting}                                                                  \\
\multicolumn{1}{c}{} & \multicolumn{1}{c}{Variable name} & \multicolumn{1}{c}{PI} & \multicolumn{1}{c}{PPE} & \multicolumn{1}{c}{Variable name} & \multicolumn{1}{c}{PI} & \multicolumn{1}{c}{PPE} & \multicolumn{1}{c}{Variable name}                   & \multicolumn{1}{c}{PI} & \multicolumn{1}{c}{PPE} \\ \hline
\multicolumn{10}{c}{\textbf{Panel A: Positive affect}}                                                   \\
1  & Age                       & .14  & -.26  & Physical health index     & .07 & .42   & Physical health index     & .16 & .62   \\
2  & Age squared               & .09  & -.26  & Learn something every day & .06 & .43   & Learn something every day & .05 & .49   \\
3  & Physical health index     & .09  & .66   & Not treated with respect  & .03 & -1.39 & Not treated with respect  & .03 & -1.13 \\
4  & Learn something every day & .05  & .82   & Health encouragement      & .02 & .13   & Health encouragement      & .02 & .14   \\
5  & Not treated with respect  & .03  & -1.52 & Diagnosed w. depression   & .01 & .27   & BMI                       & .01 & .02   \\
6  & Health encouragement      & .02  & .23   & City/area is perfect      & .00 & .17   & Diagnosed w. depression   & .01 & .34   \\
7  & In workforce              & .01  & .44   & Health limits activities  & .00 & -.01  & Has any health problems   & .01 & -.26  \\
8  & Diagnosed w. depression   & .01  & .52   & BMI                       & .00 & .09   & City/area is perfect      & .00 & .17   \\
9  & Not working               & .00  & -.32  & Age squared               & .00 & -.11  & Health limits activities  & .00 & .21   \\
10 & Tuesday                   & .00  & -.33  & Age                       & .00 & -.11  & Female                    & .00 & .20   \\ \hline
\multicolumn{10}{c}{\textbf{Panel B: Negative affect}}                                                                                                                                             \\
1  & Physical health index    & .26 & -.11 & Physical health index    & .31 & -.15 & Physical health index    & .50 & -.18 \\
2  & Not treated with respect & .03 & .16  & Not treated with respect & .04 & .17  & BMI                      & .04 & -.02 \\
3  & Diagnosed w. depression  & .02 & -.09 & BMI                      & .03 & -.01 & Not treated with respect & .03 & .15  \\
4  & Age squared              & .01 & -.03 & Diagnosed w. depression  & .02 & -.07 & Has any health problems  & .02 & .06  \\
5  & BMI                      & .01 & -.03 & Health limits activities & .01 & -.02 & Diagnosed w. depression  & .02 & -.07 \\
6  & Has any health problems  & .01 & .04  & Has any health problems  & .01 & .02  & Health limits activities & .02 & -.06 \\
7  & Cannot afford healthcare & .01 & -.05 & Cannot afford healthcare & .01 & -.04 & Had a cold yesterday     & .01 & .07  \\
8  & Wednesday                & .00 & .05  & City/area is perfect     & .00 & -.02 & Cannot afford healthcare & .01 & -.04 \\
9  & Neck or backpain         & .00 & -.03 & Neck or backpain         & .00 & -.02 & Headache yesterday       & .00 & .02  \\
10 & Time Zone E              & .00 & .03  & Age                      & .00 & -.04 & City/area is perfect     & .00 & -.02  \\\hline
\end{tabular}}
\vspace{0.1cm}

\linespread{0.9}\small
 \textbf{Notes:} The following variables are shown. Dummies: \textit{Cannot afford healthcare}, \textit{Female}, \textit{Ever diagnosed with depression},\textit{ Not treated with respect}, \textit{In workforce}, \textit{Has any health problems}, \textit{Tuesday}, \textit{Wednesday}, \textit{Neck or backpain}, \textit{Time Zone E}.  Likert scales: \textit{Learn something every day}, \textit{City/area is perfect}, \textit{Receives Health encouragement} (1 – strongly disagree, 5 – strongly agree), \textit{Economy in this country} (1 – poor to 4 – Excellent), \textit{Health limits activities in the last month} (0 to 30 days). Continuous: \textit{Age}, \textit{age squared}, \textit{Log HH income}, \textit{Physical health index}.  
\end{table}

\newpage
\begin{table}[H]
\caption{Permutation Importance (PI) and Pseudo Partial Effect (PPE) of OLS, RF and GB for positive and negative affect of the top 10 most-important variables (using 2013 SOEP data with the Extended Set of variables).}
\label{tabA6}
\resizebox{1\textwidth}{!}{\begin{tabular}{llrrlrrlrr}
\hline
\multicolumn{1}{c}{} & \multicolumn{3}{c}{OLS}                                                              & \multicolumn{3}{c}{Random forest}                                                    & \multicolumn{3}{c}{Gradient boosting}                                                                  \\
\multicolumn{1}{c}{} & \multicolumn{1}{c}{Variable name} & \multicolumn{1}{c}{PI} & \multicolumn{1}{c}{PPE} & \multicolumn{1}{c}{Variable name} & \multicolumn{1}{c}{PI} & \multicolumn{1}{c}{PPE} & \multicolumn{1}{c}{Variable name}                   & \multicolumn{1}{c}{PI} & \multicolumn{1}{c}{PPE} \\ \hline
\multicolumn{10}{c}{\textbf{Panel A: Positive affect}}                                                   \\
1  & Partner in HH                   & .03 & .21  & Partner in HH                   & .03 & .17  & Partner in HH                   & .03 & .17  \\
2  & Worry a lot                     & .02 & -.05 & Health limits social life       & .02 & .03  & Worry a lot                     & .02 & -.03 \\
3  & Health limits social life       & .02 & .08  & Number of close friends         & .01 & .07  & Health limits social life       & .01 & .05  \\
4  & Deal well with stress           & .01 & .04  & Worry a lot                     & .01 & -.02 & Number of close friends         & .01 & .08  \\
5  & Excursions/short trips          & .01 & -.07 & Deal well with stress           & .01 & .04  & Deal well with stress           & .01 & .04  \\
6  & Number of close friends         & .01 & .04  & Excursions/short trips          & .01 & -.03 & Excursions/short trips          & .01 & -.06 \\
7  & Last Word Fin. Decisions-NA     & .01 & -.08 & HH income                       & .00 & .04  & Hours of childcare per day      & .00 & .01  \\
8  & Importance: to help others      & .01 & -.07 & Attend cinema/concerts          & .00 & -.04 & Use of social networks          & .00 & -.06 \\
9  & Health limits daily life: a lot & .00 & -.12 & Am sociable                     & .00 & .01  & Importance to help others       & .00 & -.05 \\
10 & Psychiatric problems            & .00 & -.16 & Visit neighbours/friends        & .00 & -.01 & Personal patience               & .00 & .04  \\ \hline
\multicolumn{10}{c}{\textbf{Panel B: Negative affect}}                                                                                                                                             \\
1  & Worry a lot                     & .11 & .13  & Worry a lot                     & .13 & .03  & Worry a lot                     & .12 & .05  \\
2  & Health limits social life       & .04 & -.12 & Health limits social life       & .04 & -.08 & Health limits social life       & .04 & -.12 \\
3  & Female                          & .03 & .20  & Deal well with stress           & .03 & -.03 & Female                          & .02 & .16  \\
4  & Deal well with stress           & .02 & -.06 & Female                          & .02 & .15  & Deal well with stress           & .02 & -.03 \\
5  & Hours of sleep                  & .01 & -.10 & Psychiatric problems            & .01 & .18  & Number of doctor visits         & .01 & .08  \\
6  & Health limits daily life: a lot & .01 & .18  & Number of doctor visits         & .01 & .05  & Hours of sleep                  & .01 & -.07 \\
7  & Psychiatric problems            & .01 & .25  & Hours of sleep                  & .01 & -.04 & Psychiatric problems            & .01 & .22  \\
8  & Personal patience               & .01 & -.06 & Annual pension                  & .01 & .00  & Personal Patience               & .01 & -.05 \\
9  & Health affects tiring tasks     & .01 & .15  & Personal Patience               & .01 & -.03 & Annual pension                  & .01 & .00  \\
10 & Number of doctor visits         & .01 & .03  & Physical pain last 4 weeks      & .00 & -.03 & Health limits daily life: a lot & .00 & .12  \\\hline
\end{tabular}}
\vspace{0.1cm}

\linespread{0.9}\small
 \textbf{Notes:} The following variables are shown.: Dummies: \textit{Health limits daily life a lot}, \textit{Health limits daily life a bit}, \textit{Partner in HH}, \textit{No monthly savings},\textit{ Not unemployed}, \textit{No emergency reserves}, \textit{Last word in financial decisions-NA}, \textit{Psychiatric problems}, \textit{Female}, and \textit{No annual holiday trip}. Likert scales: \textit{Limited socially due to health} (1 – always to 5 – never), \textit{Worries a lot}, \textit{Importance: To help others} (1 – Very Important to 4 – Not important), \textit{Deals well with stress} (1 – not at all to 7 – totally agree), \textit{Personal patience} (0 – very bad to 10 – very good), \textit{House needs repair} (1 – in good condition, 3 – needs major renovation), \textit{Attend cinema/concerts }(1 – Daily to 4 - Infrequent), \textit{Am Sociable }(1 to 7), \textit{Visit neighbours/friends} (1 – Daily to 5 - Never), \textit{Use of social networks} (1 – Daily to 5 - Never), \textit{Health affects tiring tasks} (1 – A lot to 3 -  Not at all), and \textit{Physical pain last 4 weeks} (1 – Always to 5 - Never). Continuous: \textit{Log HH income}, \textit{Hours of sleep}, \textit{Number of doctor visits}, \textit{Risk tolerance}, \textit{Unemployment benefit}, \textit{Excursions/short trips}, \textit{Number of close friends}, \textit{Hours of childcare per day}, \textit{Annual pension}.    
\end{table}

\newpage
\begin{table}[H]
\caption{Permutation Importance (PI) of OLS, RF and GB for levels of wellbeing of the 10 most-important variables (using pooled UKHLS data with the Restricted Set of variables). For each covariate, the models include the average value and the annual deviation from that average.}
\label{tabA7}
\resizebox{1\textwidth}{!}{\begin{tabular}{llrlrlr}
\hline
 & \multicolumn{2}{c}{OLS}                       & \multicolumn{2}{c}{Random forest}             & \multicolumn{2}{c}{Gradient boosting}         \\
 & Variable name                          & PI   & Variable name                          & PI   & Variable name                          & PI   \\ \hline
1 & Health limits activities: a lot (avg.) & .041 & Age (avg.)                             & .025 & Age (avg.)                             & .026 \\
2 & Not disabled (health) (avg.)           & .020 & Not disabled (health) (avg.)           & .020 & Not disabled (health) (avg.)           & .022 \\
3 & Married (avg.)                         & .019 & Health limits activities: a lot (avg.) & .018 & Health limits activities: a lot (avg.) & .021 \\
4 & Health limits activities: a bit (avg.) & .017 & Health limits activities: a bit (avg.) & .014 & Health limits activities: a bit (avg.) & .014 \\
5 & LT sick or disabled (empl.) (avg.)     & .015 & LT sick or disabled (empl.) (avg.)     & .011 & HH income (avg.)                       & .012 \\
6 & Age (avg.)                             & .013 & HH income (avg.)                       & .009 & LT sick or disabled (empl.) (avg.)     & .012 \\
7 & Retired (avg.)                         & .012 & Married (avg.)                         & .006 & Married (avg.)                         & .009 \\
8 & HH income (avg.)                       & .010 & Retired (avg.)                         & .005 & Retired (avg.)                         & .006 \\
9 & Unemployed (avg.)                      & .007 & Unemployed (avg.)                      & .004 & Unemployed (avg.)                      & .005 \\
10 & Rents the house/flat                   & .005 & Health limits activities: a bit        & .003 & Health limits activities: a lot        & .004 \\ \hline
\hline
\end{tabular}}
\vspace{0.1cm}

\linespread{0.9}\small
 \textbf{Notes:} All covariates apart from month, ethnicity and sex are split into individual means and deviation from the mean. Individual averages are denoted by \textit{(avg.)}; variables without additional notes are the deviations from the individual means.     
\end{table}

\begin{table}[H]
\caption{Permutation Importance (PI) of OLS, RF and GB for deviations from the average wellbeing and individual level of wellbeing of the 10 most-important variables (using pooled SOEP data with the Restricted Set of variables). For each covariate, the models include the average value and the annual deviation from that average.}
\label{tabA8}
\resizebox{1\textwidth}{!}{\begin{tabular}{llrlrlr}
\hline
 & \multicolumn{2}{c}{OLS}                       & \multicolumn{2}{c}{Random forest}             & \multicolumn{2}{c}{Gradient boosting}         \\
 & Variable name                          & PI   & Variable name                          & PI   & Variable name                          & PI   \\ \hline
1 &Age (avg.)                      & .082 & Age (avg.)                     & .126 & Age (avg.)                     & .124 \\
2& Number of doctor visits (avg.)  & .039 & Adjusted Income (avg.)         & .059 & Adjusted Income (avg.)         & .049 \\
3& Adjusted Income (avg.)          & .039 & Number of doctor visits (avg.) & .041 & Number of doctor visits (avg.) & .042 \\
4& No. of children in the hh (avg.) & .025 & Not disabled (health) (avg.)   & .021 & Not disabled (health) (avg.)   & .016 \\
5& Not disabled (health) (avg.)    & .016 & No. of people in hh (avg)       & .014 & Age                            & .010 \\
6& Single (avg.)                   & .016 & No. of children in hh (avg.)    & .011 & No. of people in hh (avg.)      & .009 \\
7& Divorced (avg.)                 & .007 & House Owner                    & .009 & No. of children in hh (avg.)    & .008 \\
8& No. of people in hh (avg.)       & .006 & Age                            & .008 & Number of doctor visits        & .007 \\
9& Number of doctor visits         & .005 & Number of doctor visits        & .005 & Single                         & .006 \\
10& House Owner                     & .005 & Number of years of education   & .005 & House Owner                    & .006\\ \hline
\hline
\end{tabular}}
\vspace{0.1cm}

\linespread{0.9}\small
 \textbf{Notes:} All covariates apart from month, ethnicity and sex are split into individual means and deviation from the mean. Individual averages are denoted by \textit{(avg.)}; variables without additional notes are the deviations from the individual means.    
\end{table}

\begin{figure}[H]
\caption{The R-squared from OLS, GB and RF on the Restricted Set of variables. The R-squareds are calculated from the training data and are not representative of out-of-sample performance.}
\centering
\includegraphics[width=0.9\textwidth]{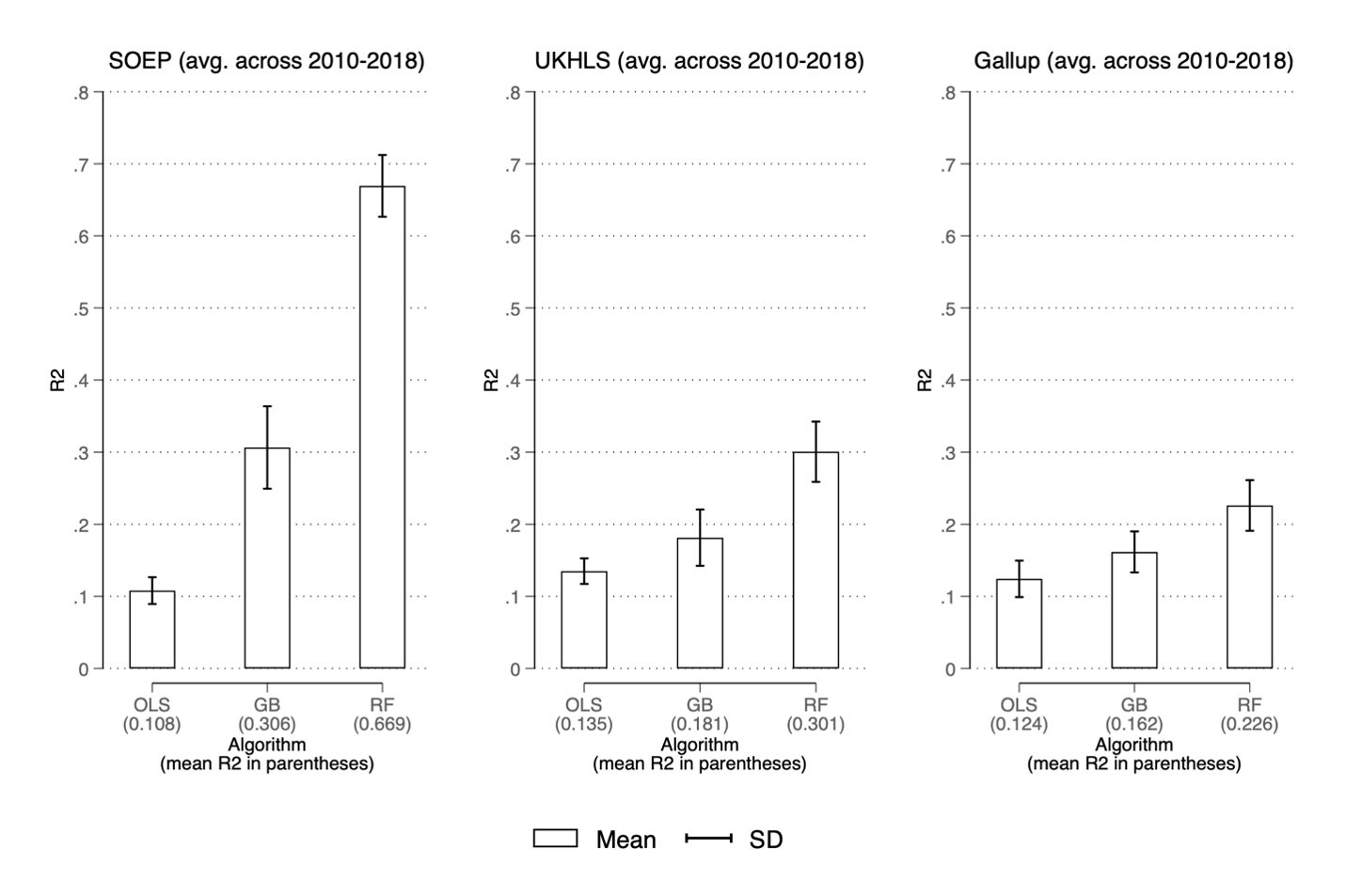}
\label{fig_a1}
\end{figure}

\begin{figure}[H]
\caption{The R-squareds from OLS, LASSO, GB, RF, and mean on the Extended Set of variables. The R-squareds are calculated from the training data and are not representative of out-of-sample performance.}
\centering
\includegraphics[width=0.9\textwidth]{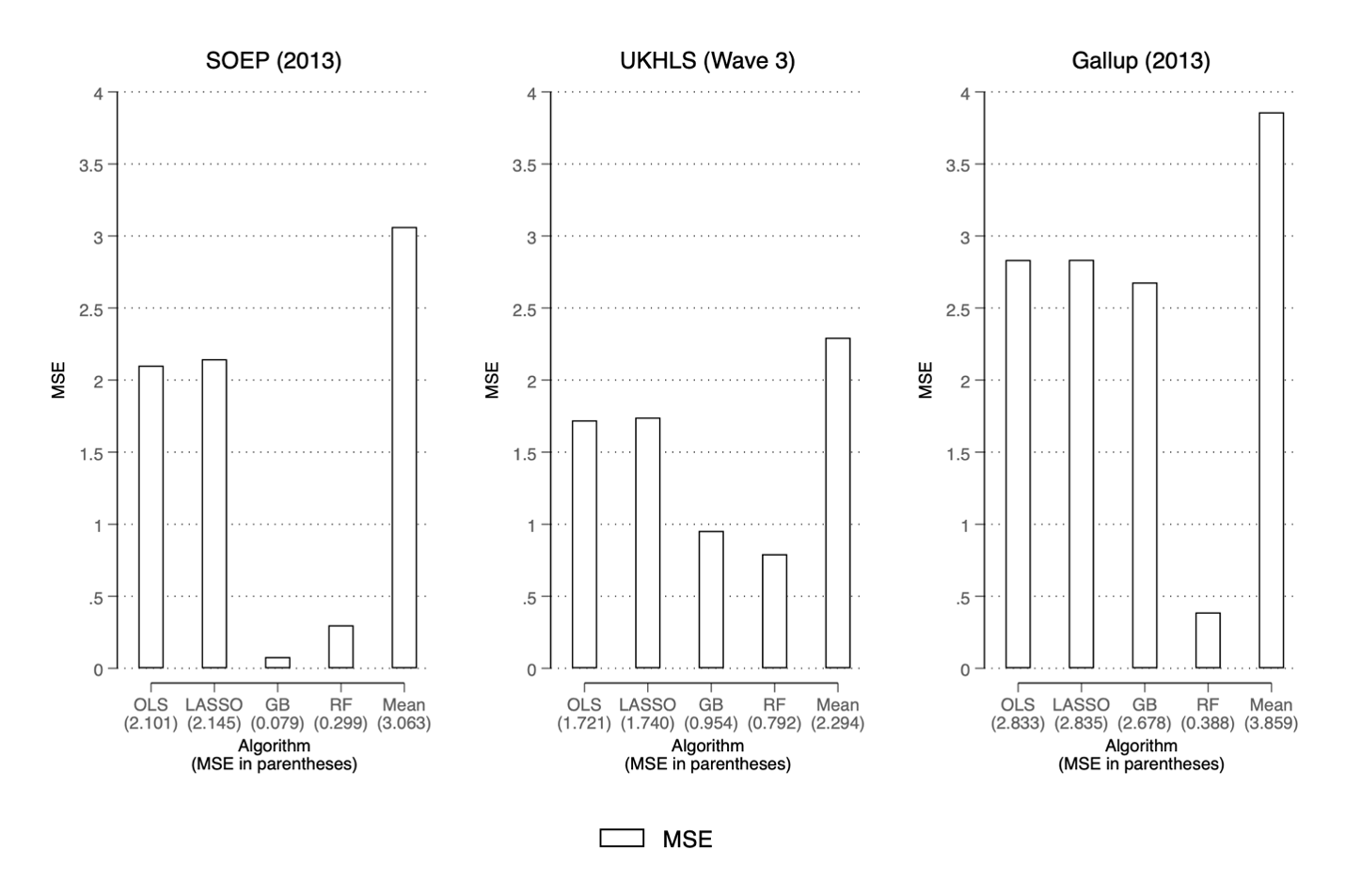}
\label{fig_a2}
\end{figure}

\begin{table}[H]
\captionof{figure}{The mean effects of age and household income on wellbeing in the Extended Set of variables.}
\label{fig_a3}
\begin{tabular}{ c c }
\multicolumn{2}{c}{\textbf{Panel A: SOEP}}     \\
\begin{minipage}{.5\textwidth}
      \includegraphics[width=\textwidth]{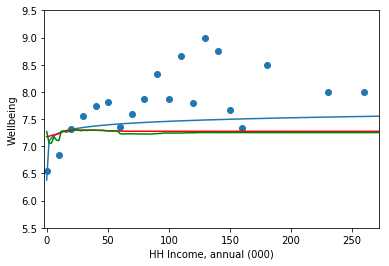}
    \end{minipage}
 & \begin{minipage}{.5\textwidth}
      \includegraphics[width=\textwidth]{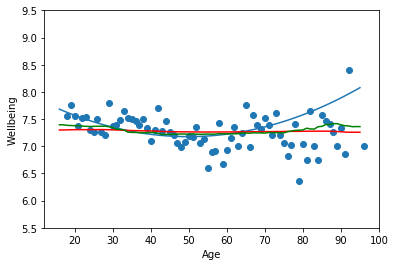}
    \end{minipage}    \\ 
\multicolumn{2}{c}{\textbf{Panel B: UKHLS}}    \\
\begin{minipage}{.5\textwidth}
      \includegraphics[width=\textwidth]{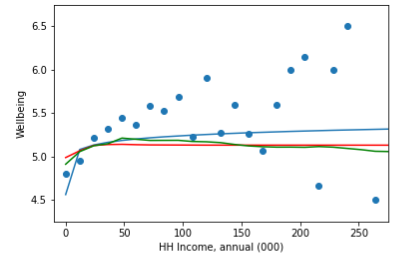}
    \end{minipage}
 & \begin{minipage}{.5\textwidth}
      \includegraphics[width=\textwidth]{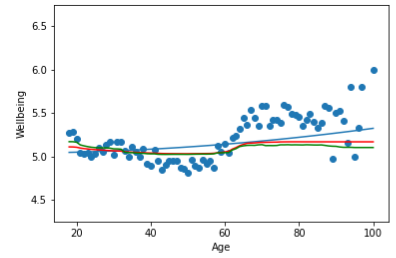}
    \end{minipage}    \\  
\multicolumn{2}{c}{\textbf{Panel C: Gallup}}   \\
\begin{minipage}{.5\textwidth}
      \includegraphics[width=\textwidth]{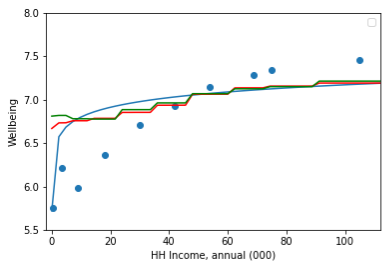}
    \end{minipage}
 & \begin{minipage}{.5\textwidth}
      \includegraphics[width=\textwidth]{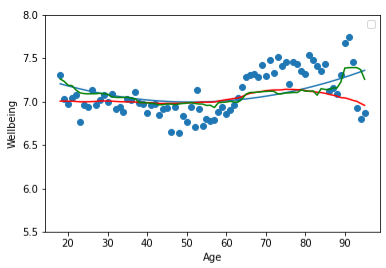}
    \end{minipage}    \\  
\multicolumn{2}{c}{\begin{minipage}{.7\textwidth}
      \includegraphics[width=\textwidth]{fig4_7.png}
    \end{minipage}}   \\    
\end{tabular}

\medskip {\small 
\textbf{Notes:} For the UKHLS and the SOEP annual income is constrained to be less than or equal to a figure of 250 000 in the local currency. This covers over 99.9\% of the income distribution in both countries. In SOEP and UKHLS, incomes are recorded as a continuous variable and equivalence-scale adjusted HH income is used for the analysis. Income data in Gallup is collected in income bands, and household size data was not collected in 2013. We here thus use non-adjusted HH income data. }
\end{table}

\begin{figure}[H]
\caption{The R-squared from OLS, LASSO, GB and RF when modelling positive and negative affect using 2013 Gallup and 2013 SOEP data with the Extended Set of variables. The R-squareds are calculated from unseen ‘testing data’.}
\centering
\includegraphics[width=0.9\textwidth]{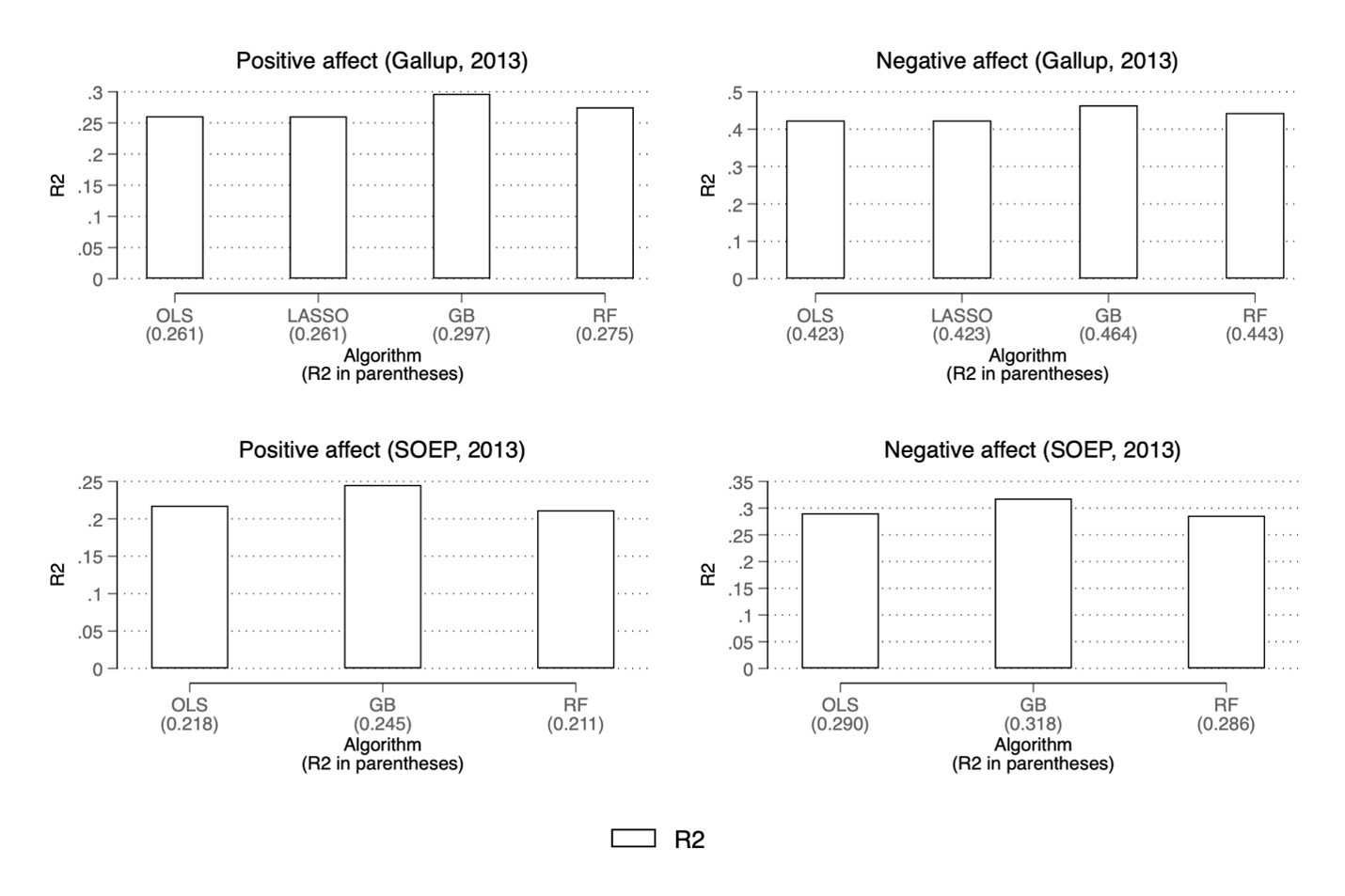}
\label{fig_a4}
\end{figure}

\begin{figure}[H]
\caption{The R-squared from OLS, GB and RF when modelling the level of wellbeing with Mundlak terms using 2013 SOEP and Wave 3 UKHLS data with the Restricted Set of variables. The R-squareds are calculated from unseen ‘testing data’.}
\centering
\includegraphics[width=0.9\textwidth]{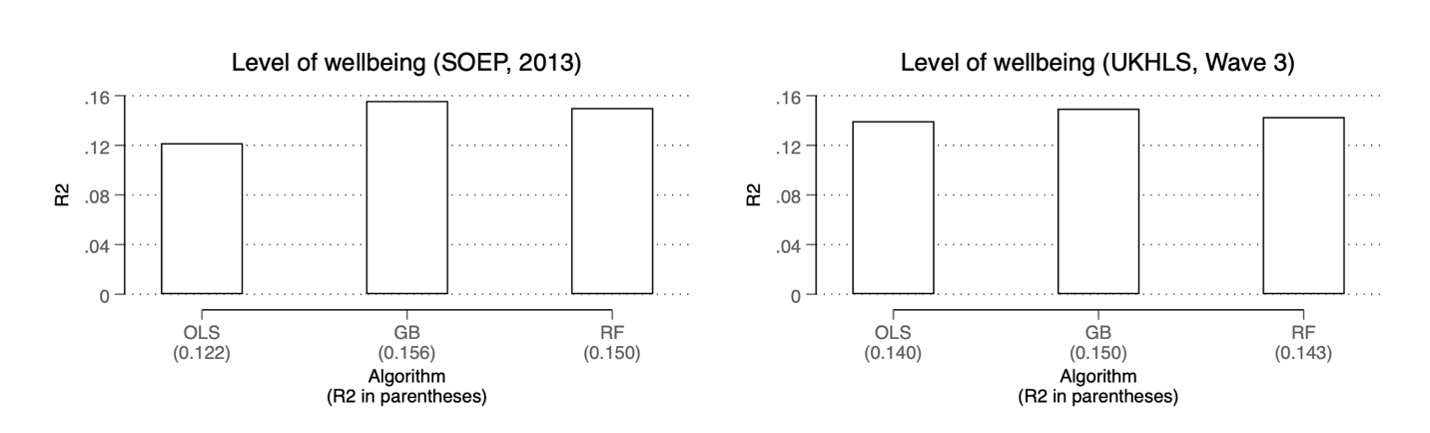}
\label{fig_a5}
\end{figure}

\end{document}